\documentclass[iop,revtex4,numberedappendix]{emulateapj}
\slugcomment{{\sc Submitted to ApJ:} November 30, 2015} 
\usepackage{graphicx}
\usepackage{natbib}
\usepackage{amssymb}
\usepackage{amsmath}
\usepackage[hyperindex,breaklinks]{hyperref}
\usepackage[figuresright]{rotating}

\newcommand{\mldyn}{$M_{\rm dyn}/L_{\mathrm{V}}$}

\def\lesssim{\lower.5ex\hbox{$\; \buildrel < \over \sim \;$}}
\def\gtrsim{\lower.5ex\hbox{$\; \buildrel > \over \sim \;$}}

\shorttitle{X-ray Emission from Ultracompact Dwarf Galaxies}
\shortauthors{Pandya, Mulchaey \& Greene}

\date{\today}

\begin{document}

\title{A Comprehensive Archival \textit{Chandra} Search for X-ray Emission from Ultracompact Dwarf Galaxies}
  
\author {Viraj Pandya$^{1,2}$, John Mulchaey$^{2}$, and Jenny E. Greene$^{1}$}
\affil{$^{1}$Department of Astrophysical Sciences, Peyton Hall, Princeton University, Princeton, NJ}
\affil{$^{2}$The Observatories of the Carnegie Institution for Science, 813 Santa Barbara Street, Pasadena, CA 91101, USA}
\email{vgpandya@princeton.edu}  

\begin{abstract}
We present the first comprehensive archival study of the X-ray properties of ultracompact dwarf (UCD) galaxies, with the goal of identifying weakly-accreting central black holes in UCDs. Our study spans 578 UCDs distributed across thirteen different host systems, including clusters, groups, fossil groups, and isolated galaxies. Of the 336 spectroscopically-confirmed UCDs with usable archival \textit{Chandra} imaging observations, 21 are X-ray-detected. Imposing a completeness limit of $L_X>2\times10^{38}$ erg s$^{-1}$, the global X-ray detection fraction for the UCD population is $\sim3\%$. Of the 21 X-ray-detected UCDs, seven show evidence of long-term X-ray time variability on the order of months to years. X-ray-detected UCDs tend to be more compact than non-X-ray-detected UCDs, and we find tentative evidence that the X-ray detection fraction increases with surface luminosity density and global stellar velocity dispersion. The X-ray emission of UCDs is fully consistent with arising from a population of low-mass X-ray binaries (LMXBs). In fact, there are fewer X-ray sources than expected using a naive extrapolation from globular clusters. Invoking the fundamental plane of black hole activity for SUCD1 near the Sombrero galaxy, for which archival \textit{Jansky Very Large Array} imaging at 5 GHz is publicly available, we set an upper limit on the mass of a hypothetical central black hole in that UCD to be $\lesssim10^5M_{\odot}$. While the majority of our sources are likely LMXBs, we cannot rule out central black holes in some UCDs based on X-rays alone, and so we address the utility of follow-up radio observations to find weakly-accreting central black holes. 
\end{abstract}

\keywords{accretion, galaxies: active, galaxies: dwarf, galaxies: supermassive black holes, X-rays: binaries, X-rays: galaxies}

\maketitle

\section{Introduction}
The formation mechanisms and central black hole occupation fraction of ultracompact dwarf (UCD) galaxies are unknown. First discovered more than fifteen years ago in the nearby Fornax cluster \citep{hilker99,drinkwater00,phillipps01}, \textit{Hubble Space Telescope} imaging and high-resolution spectroscopy have shown that UCDs are extremely dense stellar systems rather than merely unresolved foreground stars in our own Galaxy. Although numerous studies have established the distinctiveness of UCDs compared to considerably smaller and fainter globular clusters (GCs) and considerably larger and brighter dwarf elliptical (dE) galaxies \citep[e.g.,][]{drinkwater03,mieske08,norris11,brodie11}, the origin of UCDs is still not clear. There are thought to be two primary formation channels for UCDs: (1) the merging of several smaller star clusters that produces a supermassive star cluster, and (2) the tidal stripping of a larger galaxy that gives rise to a naked nucleus \citep[e.g.,][]{brodie11,norris14}.

One hint about the origin of UCDs comes from studying their dynamical masses. Interestingly, the more massive UCDs show evidence for higher dynamical mass-to-light ratios than those expected from stellar population modeling alone, suggesting that at least some UCDs host a large amount of ``dark mass" \citep{hasegan05,mieske08,mieske13}. The most prominent explanations for this dark mass include a bottom-heavy IMF \citep{mieskekroupa08}, a top-heavy IMF leading to excess stellar remnants \citep{dabringhausen12}, and supermassive black holes \citep[SMBHs;][]{mieske13}. 

Recently, \citet{seth14} presented evidence for a SMBH in M60-UCD1 that accounts for $\sim10\%$ of the total mass of that UCD, based on dynamical modeling of spatially-resolved spectroscopy. The exciting detection of a SMBH via dynamical modeling motivates us to search for alternative evidence of central black holes in UCDs, in this case using accretion. X-rays provide one of the cleanest ways to identify accretion onto SMBHs because they are insensitive to dust and can probe the low bolometric Eddington ratio regime \citep[e.g.,][]{gallo10,miller15}. 

Although the optical and UV \citep[e.g.,][]{mieske08uv} properties of UCDs have been systematically explored, X-rays from UCDs as a whole remain largely unexplored. A handful of UCD discovery papers report merely whether or not the newly-classified UCDs host an X-ray point source \citep[e.g., the first bona fide X-ray-detected UCD was SUCD1 near the Sombrero galaxy;][]{hau09}. \citet{phillipps13} present an archival study of the X-ray properties of UCDs limited to Fornax alone, finding an X-ray detection fraction of $\sim8\%$ (down to $L_X\approx6\times10^{37}$ erg s$^{-1}$). In this paper, we present the first comprehensive archival search for X-rays from UCDs in a variety of host systems, with the goal of identifying accreting central black holes. As we will show, many archival observations are deep enough to also probe low-mass X-ray binaries (LMXBs), high-mass X-ray binaries (HMXBs), and ultra-luminous X-ray sources (ULXs).

This paper is organized as follows. In \autoref{sec:sample}, we describe our sample of UCDs and their physical properties. In \autoref{sec:xrayanalysis}, we derive the X-ray properties of UCDs. In \autoref{sec:results}, we present correlations between X-ray emission and physical properties, and address the origin of the X-ray emission. We discuss our results in \autoref{sec:disc} and summarize in \autoref{sec:conc}. \autoref{sec:obs} gives the IDs of archival \textit{Chandra} datasets used, \autoref{sec:borderline} discusses borderline detections, \autoref{sec:perseus3115} comments on UCDs in Perseus and NGC 3115, and \autoref{sec:nondetprop} gives the optical properties and X-ray upper limits for non-X-ray-detected UCDs. 

\section{Sample}\label{sec:sample}
In this section, we introduce our sample of UCDs and describe their optically-derived properties.

\subsection{Optically-Defined UCDs in the Literature}
We began our study by searching the literature for all samples of optically-defined UCDs. We find a total of 578 known UCDs distributed among thirteen different host systems and four different environment types (clusters, groups, fossil groups, and isolated galaxies).\footnote{Recently, \citet{liu15b} presented a sample of several new UCDs around M87, M49, and M60 in Virgo (a small fraction of which are spectroscopically-confirmed). Although we do not include these new UCDs in our study, we do not expect our results to change if we did incorporate them.} Of those 578 UCDs, 195 are classified on the basis of their photometry alone, and so are considered UCD ``candidates." It is likely that some of these 195 UCD candidates are actually background galaxies or foreground stars. The remaining 383 UCDs are spectroscopically-confirmed members of their respective host systems, and so can be considered bona fide UCDs.\footnote{There are a few more host systems which contain mostly UCD candidates (no radial velocities) but which have no usable archival \textit{Chandra} imaging \citep[e.g., the Antlia cluster;][]{caso13}. We do not consider those systems here.} To maximize the sample size considered in this work, we consider both candidate and spectroscopically-confirmed UCDs. 

Although we do not attempt to homogenize the definition of UCDs across different studies, we have tried to limit our sample to objects that were unambiguously classified as UCDs in the original studies. The definition of an unambiguous UCD varies between different studies, and can depend on $M_V$, optical half-light radius ($r_{\rm hl}$), or a combination of the two. Typically, $M_V\lesssim-11$ mag and $r_{\rm hl}\gtrsim10$ pc for an unambiguous UCD, but there are exceptions \citep[e.g.,][]{gregg09,brodie11}. The $M_V$ distribution of our UCDs has mean $-11.26$ mag and standard deviation $0.97$ mag. The $r_{\rm hl}$ distribution of our UCDs has mean $18.2$ pc and standard deviation $14.9$ pc, with the largest value being $\sim100$ pc. Given that the boundary between GCs and faint UCDs is imprecisely defined, our sample size would greatly increase if we extended our study to encompass borderline objects \citep[e.g., the $\omega$ Centauri-like objects described in][]{hilker11}.

In \autoref{tab:refs}, we organize the list of previous studies from which we initially drew our UCD samples and properties, as a function of UCD host system. \autoref{tab:refs} also shows how many UCDs in each host system have usable archival \textit{Chandra} observations (see also \autoref{sec:data}). In addition to \autoref{tab:refs}, we also refer the reader to the following compilations of physical properties of compact stellar systems, including UCDs: \citet{norris14,norris11,forbes13,misgeldhilker11,hilker11}.

\begin{table*}
\tiny

\caption{UCD Sample\label{tab:refs}}
\begin{center}
\begin{tabular}{ccccp{12cm}}
\tableline
System & $N_{\rm tot}$ & $N_{\rm spec}$ & $N_{\rm chandra}$ & References \\
(1) & (2) & (3) & (4) & (5) \\ \tableline
Centaurus & 30 & 30 & 27 & \citet{mieske07,mieske09} \\ 
Coma & 87 & 41 & 35 & \citet{chiboucas11,madrid10} \\
Fornax & 78 & 78 & 62 & \citet{bergond07,firth08,gregg09,hilker07,mieske08,mieske13} \\
HCG 22+90 & 21 & 21 & 21 & \citet{darocha11} \\
Hydra & 83 & 54 & 67 & \citet{misgeld08,misgeld11,wehner07} \\
NGC 1023 & 15 & 0 & 1 & \citet{mieske07b} \\
NGC 1132 & 6 & 6 & 6 & \citet{madrid13,madrid11} \\
NGC 3115 & 31 & 6 & 31 & \citet{jennings14} \\
NGC 5128 & 27 & 27 & 23 & \citet{rejkuba07,mieske08,mieske13,peng04,martini04} \\
NGC 7252 & 1 & 1 & 1 & \citet{schweizer98,fellhauer05,maraston04} \\
Perseus & 84 & 14 & 28 & \citet{penny12,penny14} \\
Sombrero & 1 & 1 & 1 & \citet{hau09} \\
Virgo & 114 & 104 & 70 & \citet{brodie11,francis12,hasegan05,seth14,norris14,sandoval15,strader13,zhang15,evstigneeva08,evstigneeva07,paudel10,janz15} \\ \tableline
Total & 578 & 383 & 373 & $-$ \\
\end{tabular}
\tablecomments{References for the 13 UCD host systems probed in this paper as well as the number of UCDs in each host system satisfying our selection criteria. Col (1): UCD host system. Col (2): Total number of UCDs. Col (3): Total number of spectroscopically-confirmed UCDs. Col (4): Total number of UCDs with at least one usable Chandra observation. Col (5): References from which we drew our sample of UCDs.}
\end{center}
\end{table*}

\subsection{The Physical Properties of UCDs}
We adopt the optically-derived physical properties of our UCDs from the literature (i.e., from one of the references given in \autoref{tab:refs}). A substantial number of our UCDs do not have fundamental physical properties available. Here we briefly address the homogeneity of our literature-based physical properties.

Whenever possible, we adopt the global stellar velocity dispersion ($\sigma$) and $r_{\rm hl}$ from \citet{mieske08} and \citet{mieske13}, since they accounted for aperture- and seeing-related effects in a homogeneous way. Otherwise, we adopt the values from the original papers if available. Note that different studies used different structural models to derive sizes (Sersic, Nuker, or King profiles). In addition, we adopt the dynamical masses ($M_{\rm dyn}$), and dynamical mass-to-light ratios in the $V$ band (\mldyn), derived by \citet{mieske13} and the original papers for $\sim50$ UCDs. We verified that the $M_{\rm dyn}$ definitions used in different studies are indeed consistent with each other by recomputing the virial mass, which we estimate following \citet{spitzer87} and \citet{hilker07}: $M_{vir}\approx9.75\times\frac{r_{\rm hl}\sigma^2}{G}$.

\section{X-ray Analysis}\label{sec:xrayanalysis}
In this section, we describe all work related to deriving the X-ray properties of UCDs. 

\subsection{Chandra X-ray Data Retrieval}\label{sec:data}
For each UCD, we used the \textit{CIAO}\footnote{The Chandra Interactive Analysis of Observations software provided by the Chandra X-ray Center \citep{fruscione06}.} tasks \texttt{find\_chandra\_obsid} and \texttt{download\_chandra\_obsid} to search for archival ACIS\footnote{The Advanced CCD Imaging Spectrometer instrument.} imaging (excluding grating, continuous clocking, and interleaved mode observations) with off-axis angle\footnote{The off-axis angle is defined as the separation between the position of a source and the aim point of an observation.} $\theta<10$ arcmin. Although there was no constraint on the depth of the \textit{Chandra} data, the shortest exposure time for an individual observation is generally $5$ ksec. The restriction on $\theta$ is crucial because the \textit{Chandra} point spread function (PSF) is highly variable as a function of $\theta$ (and effective energy): the PSF starts to become elliptical at $\theta>1.0$ arcmin, and as $\theta$ increases, the radius of a circular aperture needed to enclose 90\% of the PSF also increases. If a point source has $\theta>10$ arcmin, it will look like locally-enhanced background, have an artificially-extended structure, and is unlikely to be found by most detection algorithms\footnote{See section 4.2.3 of The \textit{Chandra} Proposers' Observatory Guide: \url{http://cxc.harvard.edu/proposer/POG/html/chap4.html}}.

After all observations were downloaded, we discarded any observations whose Validation and Verification (V\&V) reports indicated severe problems. We then reprocessed every observation with the \textit{CIAO} task \texttt{chandra\_repro} to use the latest software and calibration updates (CALDB version 4.6.7). Forgoing astrometric corrections, we adopted the standard \textit{Chandra} astrometric reference frame because it has an overall 90\% absolute positional uncertainty of 0.6 arcsec, which is good enough for our purposes. 

\autoref{tab:refs} gives the number of UCDs with at least one usable \textit{Chandra} observation as a function of host system, and \autoref{sec:obs} gives the list of identifiers for all observations used in this work. We find that $373/578\approx65\%$ of UCDs have at least one usable \textit{Chandra} observation.

\subsection{Detecting X-ray Counterparts}
We take a two-tiered approach to searching for X-ray point sources at the known optical positions of UCDs. The first tier involves the use of merged exposures (individually spanning several years) to gain the maximum possible sensitivity to faint sources. The second tier involves the use of the individual exposures to account for possible variable sources that may have been missed using the merged-observation approach. 

\subsubsection{Tier I: Merged Observations}\label{sec:merged}
The process of merging \textit{Chandra} observations requires great care because of the significant degradation of the PSF at $\theta>1$ arcmin and the non-uniformity of the PSF between different observations. Within most of the thirteen UCD host systems, there are typically several different observations, some of whose aim points can be separated by a few degrees. Combining such highly-separated observations into a single merged observation can contaminate point sources (from low-$\theta$ observations) with background counts (from high-$\theta$ observations), resulting in a noisy and artificially-extended source which is unlikely to be significantly detected. Therefore, we typically created multiple merged observations for each host system by separating individual observations into groups with similar aim points.

We chose to merge observations whose aim points were within 10 arcmin of each other for two reasons: (1) that is typically the maximum $\theta$ that a bright point source can have before it looks merely like locally-enhanced background, and (2) that was the search radius used to associate archival \textit{Chandra} observations with each UCD. We accomplished this task using the machine learning $k$-means clustering algorithm which minimizes the sum of the distances from each observation's aim point to its assigned centroid position (there are $k$ centroid positions initialized by the user) via an iterative procedure. This has the advantage that each UCD is then assigned to only one of the $k$ merged images -- the image whose $k$-means centroid position is closest to the UCD's own optical position.

We used the \textit{CIAO} task \texttt{merge\_obs} to combine the individual observations associated with each of the $k$ groups, resulting in $k$ merged images. No binning of the data was done so as to maintain the native resolution of \textit{Chandra} (0.492 arcsec px$^{-1}$). For each of the merged images, we constructed a merged PSF map by doing an exposure map-weighted average of the individual contributing observations' PSF maps. The exposure map-weighted average accounts for significantly different exposure times and effective areas between the individual contributing observations.

On each of the $k$ merged images, we ran the \textit{CIAO} detection tool \texttt{wavdetect} in three different energy bands: full band (0.5-7.0 keV), soft band (0.5-2.0 keV), and hard band (2.0-7.0 keV). We used eight different wavelet scales (1, 2, 4, 8, 16, 24, 32, and 48 pixels), and required a significance threshold of 10$^{-6}$ which corresponds to one false positive detection per \textit{ACIS} CCD.

To cross-match the UCDs to the X-ray sources found by \texttt{wavdetect}, we adopted a circular matching radius of 1.5 arcsec. If a \texttt{wavdetect} X-ray source is found within 1.5 arcsec of a known UCD optical position, then that X-ray source is said to be coincident with the UCD. Although the optical radii of UCDs are typically less than one arcsec, our choice of 1.5 arcsec for the X-ray cross-matching radius helps to account for the instability and broadening of the \textit{Chandra} PSF at $\theta>1$ arcmin. We repeated this cross-matching for all three energy bands but we base our detection classifications on results in the full band. There were no instances of a source being detected in the hard and/or soft bands but not in the full band. We also did not have multiple unique X-ray sources cross-matched to a single UCD. 

We found nineteen X-ray point sources coincident with the optical positions of UCDs using this merged-observation approach. The full-band postage stamps of these nineteen X-ray point sources are shown in \autoref{fig:stamps} and their optical properties are given in \autoref{tab:optdet}. In \autoref{sec:borderline}, we discuss five borderline detections, while in \autoref{sec:perseus3115} we discuss non-detections in Perseus and NGC 3115.

\subsubsection{Tier II: Individual Observations}
While the first tier search is sensitive to very faint sources thanks to the significantly increased exposure times of merged observations, there are several instrumental and astrophysical effects that could conspire to give a non-detection in a merged image. During the first tier search, we assumed the following: (1) no intrinsic variability in a source, (2) no significant background variations among observations spanning more than ten years, and (3) no significant background contamination effects induced by merging observations whose aim points can be separated by up to 10 arcmin. 

For the above reasons, we also carried out a search on the individual (non-merged) observations. For each host system, we first ran \texttt{wavdetect} on all individual observations in the same three energy bands (full, soft, and hard) and with the same parameters mentioned previously. We then cross-matched the UCDs to the X-ray point sources found in each individual observation and looked for new matches missed during the first tier. Of course, a UCD detected in a merged image may not be detected in every individual observation contributing to that merged image because the individual observations' exposure times (and the UCD's $\theta$) may vary. In general, however, this unlocks time variability studies, which we will discuss in \autoref{sec:tvar}.

\begin{figure*}
\begin{center}
\includegraphics[width=0.9\hsize]{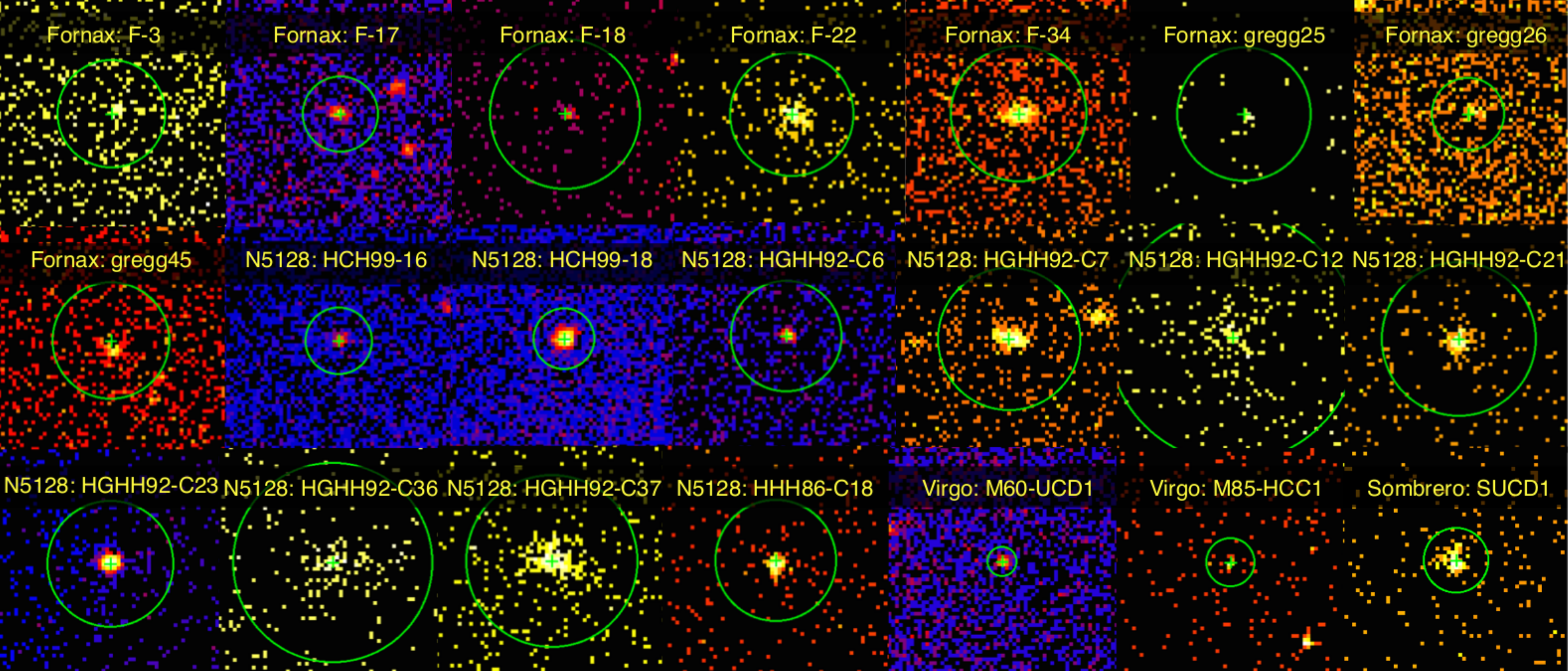}
\end{center}
\caption{Approximately $30" \times30"$ postage stamps from the full band (0.5-7.0 keV) merged image for each of the 21 X-ray-detected UCDs. The green cross marks the optical position of each UCD. The green circle is the exposure time-weighted average of the circular apertures adopted to enclose 90\% of the PSF in the individual observations (see \autoref{sec:xraylum}). The color-coding represents the count rate, where in each image blue is faintest, red is intermediate, and white is brightest.\label{fig:stamps}}
\end{figure*}

We found two additional UCDs hosting X-ray point sources using this individual-observation approach: gregg45 in Fornax and HGHH92-C12=R281 in NGC 5128. Both gregg45 and HGHH92-C12=R281 were detected in only a single exposure. Their \textit{Chandra} postage stamps are shown in \autoref{fig:stamps} and their optical properties are given in \autoref{tab:optdet}. 

\subsection{Deriving X-ray Luminosities}\label{sec:xraylum}
Since \texttt{wavdetect} is only a detection tool and not appropriate for X-ray photometry, here we describe our algorithm for deriving the flux and luminosity of each X-ray-detected UCD in the three energy bands (full, soft, and hard). There is neither a standard practice advocated by the Chandra X-ray Center nor a method widely adopted throughout the literature for deriving X-ray luminosities from a merged \textit{Chandra} dataset. It is crucial that we derive X-ray luminosities using the merged datasets because most of our detections are based on the merged images. 

For each X-ray-detected UCD, we extracted its spectrum from each of the individual observations used for its detection. We used the \textit{CIAO} task \texttt{srcflux} which automatically defines a circular source aperture centered on the optical position of each UCD with the radius chosen to enclose 90\% of the PSF (the radius depends on $\theta$ and the effective energy of the full-band: 2.3 keV). A background annulus is also automatically defined by \texttt{srcflux} with inner radius equal to the source aperture radius and outer radius equal to five times the source aperture radius.\footnote{We visually inspected the source and background apertures in every individual observation associated with each X-ray-detected UCD. In some cases, there was a bright source nearby and so we manually resized or moved the background apertures to prevent contamination.} \texttt{srcflux} automatically also applies PSF corrections to account for the fraction of counts falling outside the chosen source aperture by creating a circularly-symmetric PSF model at the UCD location using the task \texttt{arfcorr}. 

We then did an exposure-weighted sum of the individual contributing observations' source and background region spectra, using the \textit{CIAO} task \texttt{combine\_spectra}. After computing the background-subtracted spectrum\footnote{The majority of our X-ray-detected UCDs' source regions have total (i.e., including the underlying background) counts $\gg100$ so subtracting the background should be valid. In the remaining low-count cases, a standard practice advocated in the literature is to simultaneously fit the source and background spectra. However, modeling the \textit{Chandra} X-ray background is not trivial, especially with few background counts, and we do not want to introduce additional systematic uncertainties. The crucial point is that, after subtracting the background for the few low-count UCDs, we do not end up in the (non-Poissonian) negative count case.}, we assumed an absorbed power-law spectral model with the absorption fixed to the Galactic value at the optical position of the UCD \citep{stark92}. In order to determine the best-fit photon index ($\gamma$) and normalization, we followed a two-step process. 

\begin{figure*}
\begin{center}
\includegraphics[width=0.9\hsize]{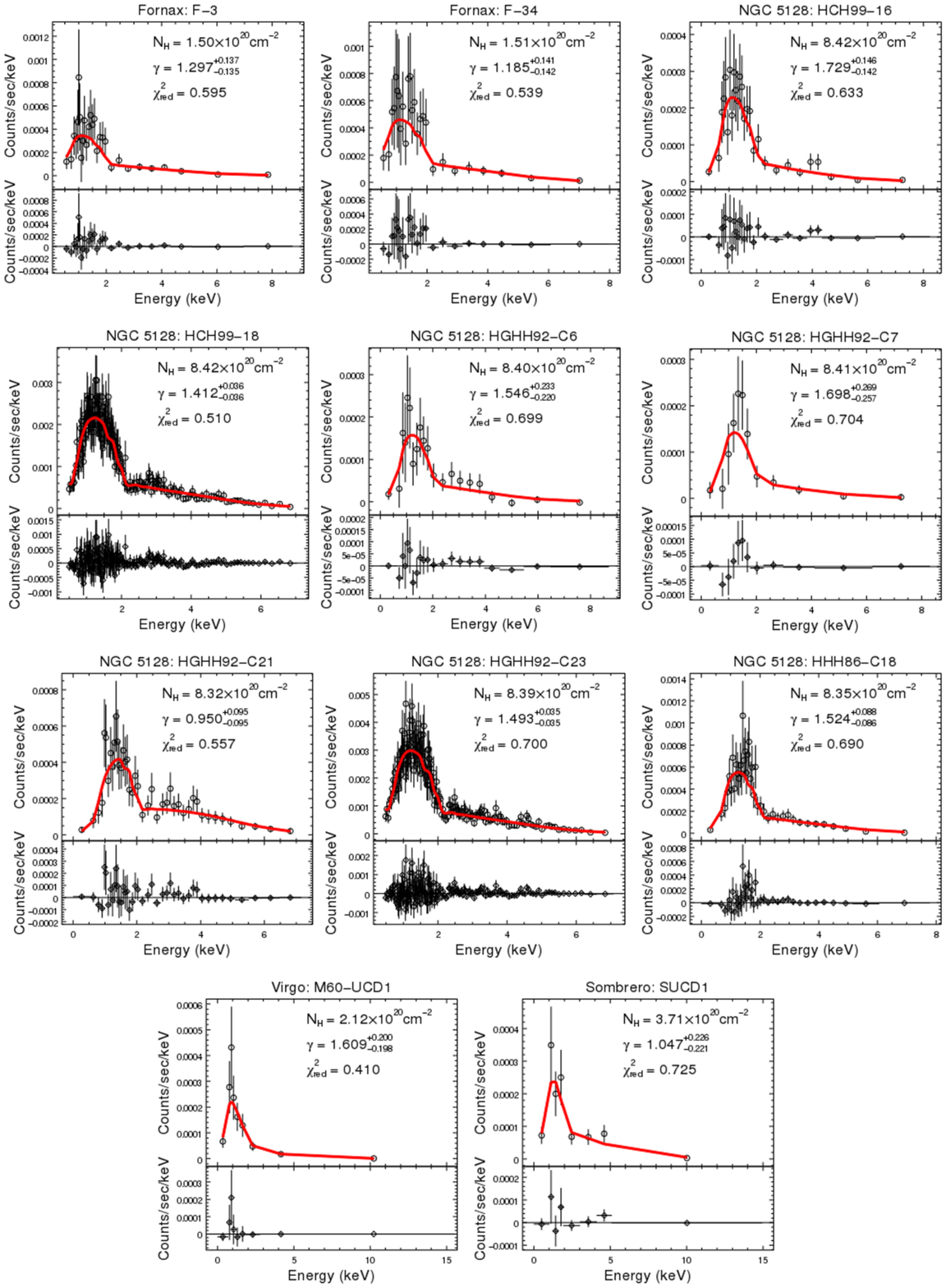}
\end{center}
\caption{Spectra and best-fit models for the subset of our X-ray-detected UCDs with $>100$ net counts. The best-fit photon index ($\gamma$) with 68\% (1$\sigma$) confidence bounds, assumed Galactic absorption value in the direction of the UCD, and reduced $\chi^2$ are shown in each subplot.\label{fig:spec}}
\end{figure*}

First, since 11 of the 21 X-ray-detected UCDs have sufficiently high net counts ($>100$ net counts) to do spectral fitting, we kept both $\gamma$ and the normalization as free parameters for these 11 UCDs. We binned the spectra to ensure at least 15 counts in each energy bin. Using the Monte Carlo fitting algorithm (\textit{moncar}) implemented in \textit{Sherpa} \citep{sherpa01} along with the \citet{gehrels86} $\chi^2$ statistic, we found reasonable spectral fits; these are shown in \autoref{fig:spec}. We investigated the reason why $\chi^2_{\rm red}<1$ for all fits as follows. For the five UCDs with $>300$ net counts, we increased the minimum bin size from 15 to 40, 80, or 100 counts and re-fit the spectra. When the error per energy bin decreased due to the larger number of counts per bin, $\chi^2_{\rm red}$ indeed increased to $\sim1$, but our fits did not significantly change. We settled on a bin size of 15 counts because some of our sources do not have enough counts to justify a bigger number and we want to apply a consistent bin size for all 11 sources. Altogether, $\chi^2_{\rm red}$ and visual inspection of the fits suggests that additional components beyond the fixed-absorption power law (e.g., soft X-ray thermal component or intrinsic absorption) are not necessarily justified.

For the remaining ten X-ray-detected UCDs, we fixed $\gamma$ to 1.5 (the median of the best-fit values derived during the first iteration), and then we solved for the best-fit normalization using the same fitting approach described above. The flux and its 90\% confidence interval in each of the three energy bands (full, soft, and hard) were derived by drawing 1000 random values from a Gaussian distribution with mean set to the best-fit model flux and standard deviation based on the uncertainty of that best-fit model flux. This was accomplished with the \textit{Sherpa} task \texttt{sample\_flux}. Finally, we converted the fluxes to luminosities with the luminosity-distance formula, assuming the distance to each UCD's host system (the distances are given in \autoref{tab:optdet}).

The X-ray properties of the X-ray-detected UCDs are given in \autoref{tab:xraydet}. 

\subsection{Deriving Upper Limits}
The majority of our UCDs are not directly detected by \texttt{wavdetect}, and so for those UCDs, we derive upper limits on the full-band (0.5-7.0 keV) flux and luminosity.\footnote{We did not derive upper limits for UCD candidates because: (1) all X-ray-detected UCDs are spectroscopically-confirmed, and (2) UCD candidates are not used when computing detection fractions or determining correlations.} In many cases, we have extraordinarily deep imaging and can place an important constraint on X-ray emission from UCDs thought to host central BHs. However, defining and deriving upper limits is a highly non-trivial and controversial topic \citep[e.g.,][]{kashyap10}. 

It is even more difficult in our case because we want to get the deepest possible upper limits by leveraging the summed exposure times of merged observations, but the individual contributing observations each have different effective exposures and aim points, and can span across more than 10 years. Tools such as WebPIMMS\footnote{See \url{https://heasarc.gsfc.nasa.gov/docs/software/tools/pimms.html}.} are not appropriate because they make two assumptions that are generally flawed for large archival studies such as ours: (1) the source is perfectly on-axis, and (2) the sensitivity of ACIS for only one observing cycle is used. In practice, the source is not perfectly on-axis ($\theta>0$ arcmin), and merged observations comprise datasets taken in many different observing cycles. Due to these complications, and because there is no standard algorithm recommended for deriving upper limits based on merged \textit{Chandra} datasets, we describe our method in \autoref{sec:nondetprop}. In short, we derived upper limits for the non-X-ray-detected UCDs by computing the number of net counts that would correspond to a local background fluctuation with Poisson probability $1\%$.

The optical properties and X-ray upper limits for non-X-ray-detected UCDs are given in \autoref{sec:nondetprop}.

\subsection{X-ray Time Variability}\label{sec:tvar}
Many of our UCDs are covered in multiple archival observations spanning nearly fifteen years, and here we take advantage of that fact to search for time variability using the data products created during the X-ray luminosity derivation process (see \autoref{sec:xraylum}). We restrict the following time variability analysis only to X-ray-detected UCDs because we know that: (1) they are definitely X-ray emitters, and (2) they are securely detected in at least one archival observation. Individual archival observations in which an otherwise X-ray-detected UCD is not ``re-detected" by \texttt{wavdetect} are still included in our analysis to test the possibility that the UCD's X-ray source ``turned off." However, we discard any observations in which \texttt{srcflux} can only measure a 90\% upper limit instead of a net count rate with 90\% confidence bounds; this is typically a problem only for observations with exposure times $\lesssim10$ ksec. 

As is standard practice in active galactic nucleus (AGN) variability studies \citep[e.g.,][]{nandra97,ptak98,vaughan03,thornton09,young12}, we searched for variability in net count rate space rather than flux or luminosity space. Following \citet{luo13}, we took a two-pronged approach to test for long-term X-ray time variability. First, we fit each UCD's ``X-ray light curve" with a zero-slope straight-line model and computed $\chi^2_{\rm red}$. Second, we computed the maximal time variability indicator, $\sigma_{\rm max}$, based on the maximum difference between any two net count rates within a given UCD's light curve, normalized by their uncertainties \citep[again, see][and references therein]{luo13}:

\begin{equation}\label{eq:tvar}
\sigma_{\rm max} = \mathrm{max_{i,j}} \frac{| R_i - R_j |}{\sqrt{\sigma_{R_i}^2+\sigma_{R_j}^2}} \; ,
\end{equation}
where $R_i$ and $R_j$ are the net count rates in the $i$th and $j$th observations, and $\sigma_{R_i}$ and $\sigma_{R_j}$ are the uncertainties in those net count rates, respectively. The uncertainties $\sigma_R$ include the contribution from the background and were derived using the \citet{gehrels86} approximation.

\begin{figure}
\begin{center}
\includegraphics[width=0.95\hsize]{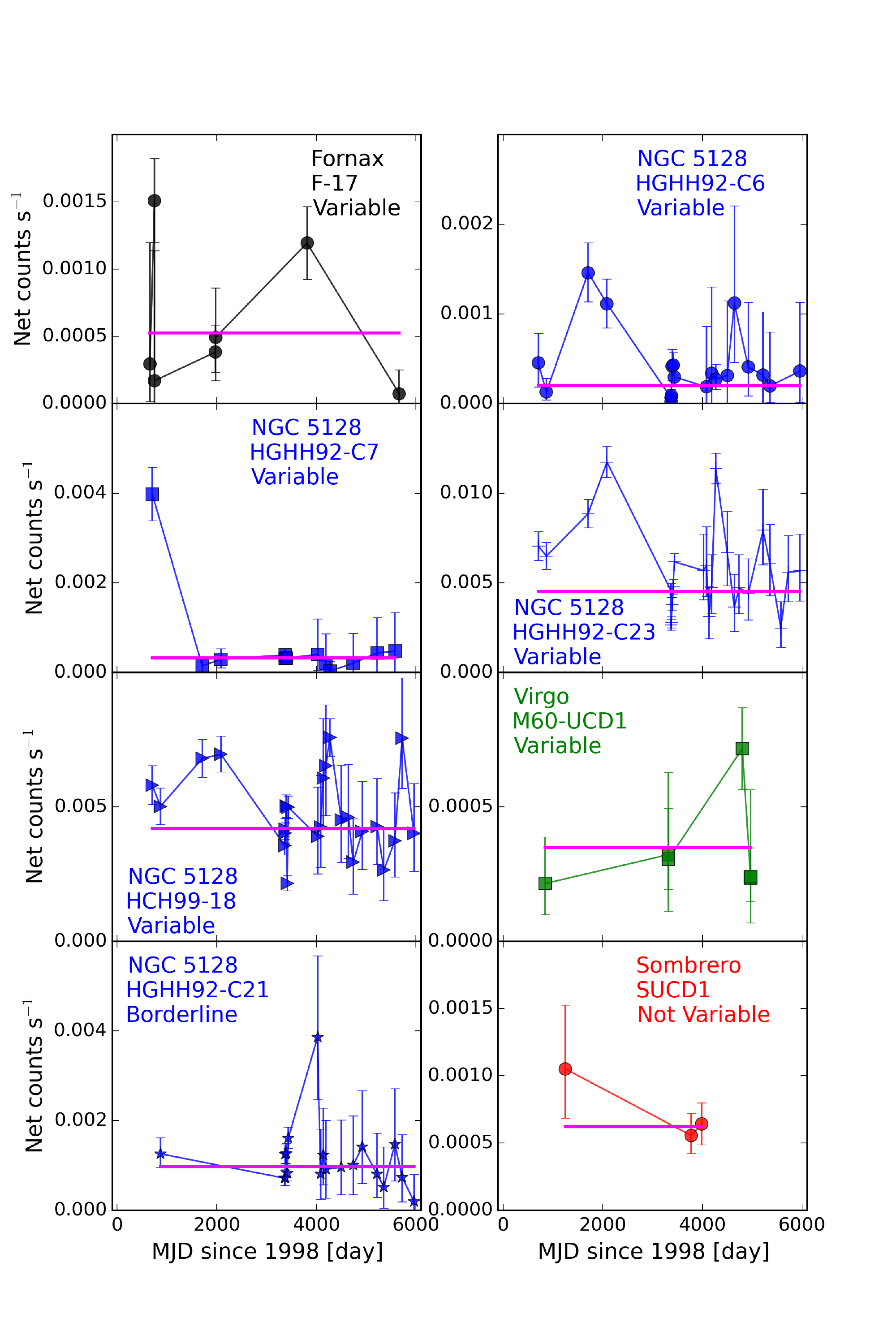}
\end{center}
\caption{The long-term X-ray light curves for the seven variability candidates and the non-variable SUCD1. The modified Julian dates (MJD) shown are relative to the \textit{Chandra}-specific reference MJD: MJD$_{ref}=50814$ days (January 1, 1998).\label{fig:tvar}}
\end{figure}

We present the results of the two approaches in \autoref{tab:tvar}. Adopting slightly more restrictive criteria than \citet{luo13}, we consider a UCD to be a variability candidate if $\chi^2_{\rm red}\gtrsim2$ and $\sigma_{\rm max}\gtrsim3$. The following five UCDs clearly satisfy both criteria: F-17, HGHH92-C6, HGHH92-C7, HGHH92-C23, and HCH99-18. M60-UCD1 also exhibits evidence for variability, albeit with somewhat lower values of $\chi^2_{\rm red}$ and $\sigma_{\rm max}$. HGHH92-C21 is a borderline case due to a single spike in net count rate. The data suggest long-term variability on the order of months to years for these seven UCDs, and their median $\sigma_{\rm max}\approx5.3$. Many of the non-variable UCDs have fewer observations available for characterizing their long-term variability as compared to the variable sources (see the $N_{\rm obs}$ column in \autoref{tab:tvar}), but their observations do span several years (with the exception of $\sim3$ months for gregg25). The long-term X-ray light curves of the seven variability candidates and the non-variable SUCD1 are shown in \autoref{fig:tvar}. 

Lastly, we also looked for short-term variability (within a single observation -- on the order of ksec) from each of the X-ray-detected UCDs using only observations in which the source region yielded $>50$ net counts. First, we used the CIAO task \texttt{dither\_region} to account for any time periods where our sources of interest were dithered over bad regions (off-chip, bad pixel, etc.) since that would induce artificial time variability due to instrumental effects. Then, we employed the \textit{CIAO} task \texttt{glvary} which uses the \citet{glvary92} algorithm to: (1) slice an observation's event file into different time bins, and (2) look for significant deviations between the time-binned events. We did not find any compelling evidence for short-term time variability within any of the UCDs' observations.

Ideally, we would like to derive a characteristic long-term variability timescale and relate that to a hypothetical black hole mass. However, we do not have a sufficient number of data points in the light curves to detect a characteristic long-term variability timescale or amplitude. We also do not detect compelling evidence for short-term variability and thus cannot derive an estimate of, e.g., the crossing time, which directly probes the black hole mass \citep[e.g., see section 2 of][]{peterson01}. Nevertheless, the long-term variability of F-17, HGHH92-C6, HGHH92-C7, HGHH92-C21, HGHH92-C23, HCH99-18, and M60-UCD1 warrants further investigation.

\begin{table*}
\begin{center}

\caption{Long-term X-ray Time Variability\label{tab:tvar}}
\small
\begin{tabular}{cccccc}
\tableline
UCD & System & N$_{\rm obs}$ & $\sigma_{\rm max}$ & $\chi^2_{\rm red}$ & Variable? \\ 
(1) & (2) & (3) & (4) & (5) & (6) \\ \tableline
F-17 & Fornax & 7 & 3.99 & 3.86 & Yes \\
F-22 & Fornax & 4 & 1.20 & 0.56 & No \\
F-34 & Fornax & 9 & 1.47 & 0.34 & No \\
F-3 & Fornax & 3 & 0.88 & 0.40 & No \\
F-18 & Fornax & 9 & 1.05 & 0.21 & No \\
gregg25 & Fornax & 2 & 0.58 & 0.33 & No \\
gregg26 & Fornax & 4 & 0.84 & 0.32 & No \\
HGHH92-C6 & NGC 5128 & 19 & 4.16 & 2.22 & Yes \\
HGHH92-C7 & NGC 5128 & 12 & 6.22 & 3.69 & Yes \\
HHH86-C18 & NGC 5128 & 21 & 1.77 & 0.51 & No \\
HGHH92-C21 & NGC 5128 & 19 & 3.04 & 1.16 & Maybe \\
HGHH92-C23 & NGC 5128 & 24 & 9.82 & 11.79 & Yes \\
HGHH92-C36=R113 & NGC 5128 & 4 & 0.24 & 0.03 & No \\
HGHH92-C37=R116 & NGC 5128 & 5 & 0.94 & 0.31 & No \\
HCH99-16 & NGC 5128 & 23 & 1.16 & 0.27 & No \\
HCH99-18 & NGC 5128 & 24 & 7.17 & 5.68 & Yes \\
SUCD1 & Sombrero & 3 & 0.99 & 0.50 & No \\
M60-UCD1 & Virgo & 6 & 2.54 & 1.52 & Yes \\
HGHH92-C12=R281 & NGC 5128 & 6 & 0.68 & 0.16 & No \\
gregg45 & Fornax & 8 & 0.98 & 0.38 & No \\
\end{tabular}
\tablecomments{The long-term X-ray time variability properties of X-ray-detected UCDs with multiple observations. Col (1): UCD name. Col (2): UCD host system. Col (3): Number of individual observations in which \texttt{srcflux} could measure a net count rate with $90\%$ confidence bounds (regardless of detection). Col (4): The maximal time variability indicator. Col (5): The reduced chi-squared from assuming a zero-slope straight-line fit model. Col (6): Whether we consider the UCD to exhibit long-term X-ray time variability.}
\end{center}
\end{table*}

\begin{sidewaystable*}
\caption{Optical Properties of X-ray-detected UCDs\label{tab:optdet}}
\tiny
\resizebox{\linewidth}{!}{
\begin{tabular}{ccccccccccccc}
\tableline
UCD & System & RA & Dec & $D$ & $M_V$ & log(M/M$_{\odot}$) & r$_{\rm hl}$ & $v_{\rm rad}$ & $\sigma$ & \mldyn  & $n_{\rm LMXB}$ & Ref \\
(1) & (2) & (3) & (4) & (5) & (6) & (7) & (8) & (9) & (10) & (11) & (12) & (13) \\\tableline
F-17 & Fornax & 54.64083 & $-$35.43255 & 20.9 & $-$11.37 & 6.8$\pm$5.8 & 3.46 & 1390.0 & 28.0$\pm$1.4 & 2.1$\pm$0.4 & 0.901 & Mieske+13 \\
F-22 & Fornax & 54.82375 & $-$35.42503 & 20.9 & $-$11.22 & 6.9$\pm$5.9 & 5.34 & 1054.0 & 29.1$\pm$1.5 & 3.2$\pm$0.9 & 0.290 & Mieske+13 \\
F-34 & Fornax & 54.55291 & $-$35.48250 & 20.9 & $-$10.83 & 6.3$\pm$5.6 & 4.19 & 1639.0 & 15.3$\pm$1.5 & 1.2$\pm$0.4 & 0.318 & Mieske+13 \\
F-3 & Fornax & 54.54333 & $-$35.40172 & 20.9 & $-$11.70 & $-$ & $-$ & 1626.0 & 31.3$\pm$1.5 & $-$ & $-$ & Mieske+08a \\
F-18 & Fornax & 54.67500 & $-$35.55361 & 20.9 & $-$11.70 & $-$ & $-$ & 2024.0 & 19.1$\pm$1.4 & $-$ & $-$ & Mieske+08a \\
gregg25 & Fornax & 54.55008 & $-$35.66589 & 20.9 & $-$ & $-$ & $-$ & 1307.0 & $-$ & $-$ & $-$ & Gregg+09 \\
gregg26 & Fornax & 54.55937 & $-$35.44550 & 20.9 & $-$ & $-$ & $-$ & 1377.0 & $-$ & $-$ & $-$ & Gregg+09 \\
gregg45$^*$ & Fornax & 54.67975 & $-$35.46708 & 20.9 & $-$ & $-$ & $-$ & 1574.0 & $-$ & $-$  & $-$ & Gregg+09 \\
HGHH92-C6 & NGC 5128 & 201.34245 & $-$43.04600 & 3.42 & $-$11.01 & 6.5$\pm$5.7 & 4.3 & 854.5 & 20.7$\pm$1.5 & 1.5$\pm$0.4 & 0.369 & Mieske+13 \\
HGHH92-C7 & NGC 5128 & 201.52254 & $-$42.94233 & 3.42 & $-$11.09 & 6.8$\pm$6.3 & 7.5 & 594.9 & 19.1 & 2.7 & 0.118 & Mieske+08a \\
HHH86-C18 & NGC 5128 & 201.41616 & $-$43.08386 & 3.42 & $-$10.14 & 6.2$\pm$5.7 & 3.2 & 479.8 & 14.7 & 0.9$\pm$0.3 & 0.264 & Mieske+08a \\
HGHH92-C21 & NGC 5128 & 201.46975 & $-$43.09622 & 3.42 & $-$10.39 & 6.7$\pm$6.2 & 7.0 & 461.3 & 17.2 & 3.9$\pm$1.4 & 0.062 & Mieske+08a \\
HGHH92-C23 & NGC 5128 & 201.47741 & $-$42.99039 & 3.42 & $-$11.66 & 6.8$\pm$6.3 & 3.3 & 673.7 & 29.5 & 1.7$\pm$0.6 & 1.393 & Mieske+08a \\
HGHH92-C36=R113 & NGC 5128 & 201.53220 & $-$42.86675 & 3.42 & $-$9.91 & 6.3$\pm$5.8 & 3.6 & 702.7 & 14.7 & 2.2$\pm$0.8 & 0.156 & Mieske+08a \\
HGHH92-C37=R116 & NGC 5128 & 201.54408 & $-$42.89519 & 3.42 & $-$9.83 & 6.0$\pm$5.6 & 3.3 & 611.7 & 12.0 & 1.5$\pm$0.5 & 0.173 & Mieske+08a \\
HCH99-16 & NGC 5128 & 201.37620 & $-$42.99300 & 3.42 & $-$10.08 & 6.3$\pm$5.8 & 12.1 & 458.2 & 8.4 & 2.2$\pm$0.8 & 0.013 & Mieske+08a \\
HCH99-18 & NGC 5128 & 201.38167 & $-$43.00078 & 3.42 & $-$11.38 & 7.0$\pm$6.6 & 13.7 & 455.0 & 18.7 & 3.7$\pm$1.4 & 0.043 & Mieske+08a \\
HGHH92-C12=R281$^*$ & NGC 5128 & 201.27383 & $-$43.17519 & 3.42 & $-$10.52 & 6.63 & 10.8$\pm$1.4 & 440.4 & 13.1 & 3.1 & 0.027 & Rejkuba+07 \\
SUCD1 & Sombrero & 190.01304 & $-$11.66786 & 9.95 & $-$12.31 & 7.3 & 14.7$\pm$1.4 & 1293.1 & 25.0$\pm$5.6 & 3.0 & 0.106 & Hau+09 \\
M60-UCD1 & Virgo & 190.89990 & 11.53470 & 16.7 & $-$14.20 & 8.3$\pm$7.5 & 24.2$\pm$0.5 & 1290.0 & 68$\pm$5 & 4.9$\pm$0.7 & 0.302 & Strader+13 \\
M85-HCC1 & Virgo & 186.34516 & 18.18158 & 16.7 & $-$12.55 & $-$ & 1.85$\pm$0.9 & 658.0 & $-$ & $-$ & 13.875 & Sandoval+15 \\
\end{tabular}}
\textbf{Notes:} Col (1): UCD name adopted from the literature. $*$ means that the UCD was discovered during the second tier search using individual rather than merged observations. Col (2): UCD host system. Col (3): Optical right ascension of UCD [deg]. Col (4): Optical declination of UCD [deg]. Col (5): Distance adopted to UCD [Mpc]. Col (6): Absolute $V$-band magnitude [mag]. Col (7): $\log_{10}$ of (dynamical mass)/$M_{\odot}$. Col (8): Optical half-light radius [pc]. Col (9): Heliocentric radial velocity [km s$^{-1}$]. Col (10): Global stellar velocity dispersion [km s$^{-1}$]. Col (11): Dynamical mass-to-light ratio in the $V$-band [$M_{\odot}/L_{\odot}$]. Col (12): Expected number of LMXBs according to equation (17) of \citet{sivakoff07}. Col (13): Primary reference for the UCD's optical properties. 
\end{sidewaystable*}

\begin{sidewaystable*}
\caption{X-ray Properties of X-ray-detected UCDs\label{tab:xraydet}}
\tiny
\resizebox{\linewidth}{!}{
\begin{tabular}{ccccccccccccccccc}
\tableline
UCD & System & EXP & $\theta$ & $s_x$ & $R$ & NC$_F$ & NC$_H$ & NC$_S$ & F$_F$ & F$_H$ & F$_S$ & L$_F$ & L$_H$ & L$_S$ & HR & $\gamma$ \\
(1) & (2) & (3) & (4) & (5) & (6) & (7) & (8) & (9) & (10) & (11) & (12) & (13) & (14) & (15) & (16) & (17) \\\tableline
F-17 & Fornax & 240 & 3.6 & 0.37 & 4.9 & 169.61$\pm$44.49 & 26.16$\pm$35.66 & 144.36$\pm$26.71 & $-$14.30$^{+0.17}_{-0.27}$ & $-$14.44$^{+0.35}_{-0.76}$ & $-$14.75$^{+0.17}_{-0.30}$ & 38.42$^{+0.17}_{-0.27}$ & 38.27$^{+0.35}_{-0.76}$ & 37.97$^{+0.17}_{-0.30}$ & $-$0.69 & 1.49 \\
F-22 & Fornax & 93 & 5.3 & 0.59 & 8.0 & 126.79$\pm$42.60 & 29.12$\pm$35.56 & 99.67$\pm$23.60 & $-$14.05$^{+0.20}_{-0.35}$ & $-$14.11$^{+0.36}_{-0.94}$ & $-$14.50$^{+0.18}_{-0.35}$ & 38.67$^{+0.20}_{-0.35}$ & 38.61$^{+0.36}_{-0.94}$ & 38.22$^{+0.18}_{-0.35}$ & $-$0.55 & 1.49 \\
F-34 & Fornax & 279 & 5.3 & 0.13 & 8.8 & 272.86$\pm$47.47 & 86.16$\pm$37.62 & 186.37$\pm$29.02 & $-$13.98$^{+0.12}_{-0.17}$ & $-$14.11$^{+0.21}_{-0.48}$ & $-$14.55$^{+0.13}_{-0.19}$ & 38.74$^{+0.12}_{-0.17}$ & 38.61$^{+0.21}_{-0.48}$ & 38.17$^{+0.13}_{-0.19}$ & $-$0.37 & 1.19$^{+0.14}_{-0.14}$ \\
F-3 & Fornax & 414 & 4.5 & 0.89 & 7.0 & 288.53$\pm$47.85 & 92.87$\pm$37.84 & 195.32$\pm$29.36 & $-$14.15$^{+0.11}_{-0.16}$ & $-$14.29$^{+0.21}_{-0.49}$ & $-$14.68$^{+0.13}_{-0.18}$ & 38.56$^{+0.11}_{-0.16}$ & 38.42$^{+0.21}_{-0.49}$ & 38.04$^{+0.13}_{-0.18}$ & $-$0.36 & 1.30$^{+0.14}_{-0.14}$ \\
F-18$^\dagger$ & Fornax & 204 & 5.5 & 0.51 & 9.7 & 32.87$\pm$43.22 & 0.96$\pm$36.98 & 31.92$\pm$22.46 & $-$14.68$^{+0.36}_{-0.87}$ & $-$14.37$^{+0.43}_{-1.00}$ & $-$15.11$^{+0.39}_{-0.98}$ & 38.04$^{+0.36}_{-0.87}$ & 38.35$^{+0.43}_{-1.00}$ & 37.61$^{+0.39}_{-0.98}$ & $-$0.94 & 1.49 \\
gregg25$^\dagger$ & Fornax & 74 & 4.7 & 0.82 & 8.7 & 18.12$\pm$40.12 & 8.00$\pm$35.00 & 10.12$\pm$19.69 & $-$14.49$^{+0.41}_{-1.04}$ & $-$14.05$^{+0.42}_{-0.95}$ & $-$14.92$^{+0.39}_{-0.91}$ & 38.22$^{+0.41}_{-1.04}$ & 38.67$^{+0.42}_{-0.95}$ & 37.80$^{+0.39}_{-0.91}$ & $-$0.12 & 1.49 \\
gregg26$^\dagger$ & Fornax & 194 & 2.7 & 1.21 & 4.7 & 5.99$\pm$40.20 & 1.00$\pm$34.78 & 4.91$\pm$20.25 & $-$15.31$^{+0.44}_{-1.05}$ & $-$14.58$^{+0.44}_{-1.01}$ & $-$15.74$^{+0.44}_{-1.02}$ & 37.41$^{+0.44}_{-1.05}$ & 38.14$^{+0.44}_{-1.01}$ & 36.97$^{+0.44}_{-1.02}$ & $-$0.66 & 1.49 \\
gregg45$^*$$^\dagger$ & Fornax & 265 & 4.8 & 1.49 & 7.6 & 45.20$\pm$41.89 & 4.00$\pm$35.17 & 41.04$\pm$22.83 & $-$14.87$^{+0.34}_{-0.93}$ & $-$14.70$^{+0.42}_{-1.04}$ & $-$15.31$^{+0.39}_{-0.89}$ & 37.85$^{+0.34}_{-0.93}$ & 38.02$^{+0.42}_{-1.04}$ & 37.41$^{+0.39}_{-0.89}$ & $-$0.82 & 1.49 \\
HGHH92-C6 & NGC 5128 & 612 & 4.4 & 0.35 & 7.2 & 175.11$\pm$46.02 & 67.95$\pm$38.24 & 108.00$\pm$25.70 & $-$14.61$^{+0.18}_{-0.29}$ & $-$14.72$^{+0.31}_{-0.86}$ & $-$15.09$^{+0.19}_{-0.35}$ & 36.53$^{+0.18}_{-0.29}$ & 36.42$^{+0.31}_{-0.86}$ & 36.06$^{+0.19}_{-0.35}$ & $-$0.23 & 1.55$^{+0.23}_{-0.22}$ \\
HGHH92-C7 & NGC 5128 & 445 & 5.4 & 0.45 & 9.2 & 106.16$\pm$43.32 & 30.37$\pm$36.21 & 76.71$\pm$23.89 & $-$14.70$^{+0.24}_{-0.59}$ & $-$14.72$^{+0.37}_{-1.05}$ & $-$15.13$^{+0.25}_{-0.62}$ & 36.45$^{+0.24}_{-0.59}$ & 36.42$^{+0.37}_{-1.05}$ & 36.02$^{+0.25}_{-0.62}$ & $-$0.43 & 1.70$^{+0.27}_{-0.26}$ \\
HHH86-C18 & NGC 5128 & 593 & 5.0 & 0.57 & 7.9 & 597.54$\pm$51.70 & 210.46$\pm$40.53 & 388.96$\pm$32.19 & $-$13.91$^{+0.06}_{-0.07}$ & $-$14.08$^{+0.12}_{-0.16}$ & $-$14.40$^{+0.06}_{-0.08}$ & 37.23$^{+0.06}_{-0.07}$ & 37.06$^{+0.12}_{-0.16}$ & 36.75$^{+0.06}_{-0.08}$ & $-$0.3 & 1.52$^{+0.09}_{-0.09}$ \\
HGHH92-C21 & NGC 5128 & 639 & 6.1 & 0.26 & 10.0 & 602.25$\pm$52.19 & 311.54$\pm$42.82 & 293.46$\pm$29.98 & $-$13.86$^{+0.06}_{-0.07}$ & $-$13.95$^{+0.10}_{-0.12}$ & $-$14.58$^{+0.08}_{-0.09}$ & 37.29$^{+0.06}_{-0.07}$ & 37.20$^{+0.10}_{-0.12}$ & 36.57$^{+0.08}_{-0.09}$ & 0.03 & 0.95$^{+0.10}_{-0.10}$ \\
HGHH92-C23 & NGC 5128 & 652 & 5.1 & 0.20 & 8.2 & 3586.70$\pm$82.61 & 1312.04$\pm$57.91 & 2286.62$\pm$59.09 & $-$13.15$^{+0.02}_{-0.02}$ & $-$13.31$^{+0.03}_{-0.03}$ & $-$13.65$^{+0.02}_{-0.02}$ & 38.00$^{+0.02}_{-0.02}$ & 37.83$^{+0.03}_{-0.03}$ & 37.50$^{+0.02}_{-0.02}$ & $-$0.27 & 1.49$^{+0.04}_{-0.04}$ \\
HGHH92-C36=R113 & NGC 5128 & 106 & 7.1 & $-$ & 12.9 & 69.42$\pm$41.41 & 5.67$\pm$35.09 & 64.75$\pm$22.12 & $-$14.05$^{+0.25}_{-0.61}$ & $-$14.10$^{+0.39}_{-0.94}$ & $-$14.56$^{+0.27}_{-0.68}$ & 37.09$^{+0.25}_{-0.61}$ & 37.05$^{+0.39}_{-0.94}$ & 36.59$^{+0.27}_{-0.68}$ & $-$0.84 & 1.49 \\
HGHH92-C37=R116 & NGC 5128 & 291 & 6.5 & $-$ & 11.1 & 124.92$\pm$42.96 & 35.33$\pm$36.07 & 89.46$\pm$23.40 & $-$14.28$^{+0.18}_{-0.37}$ & $-$14.39$^{+0.32}_{-0.76}$ & $-$14.79$^{+0.20}_{-0.40}$ & 36.86$^{+0.18}_{-0.37}$ & 36.75$^{+0.32}_{-0.76}$ & 36.36$^{+0.20}_{-0.40}$ & $-$0.43 & 1.49 \\
HCH99-16 & NGC 5128 & 837 & 3.4 & 0.32 & 4.3 & 324.79$\pm$47.66 & 100.02$\pm$38.21 & 226.65$\pm$28.61 & $-$14.46$^{+0.09}_{-0.12}$ & $-$14.68$^{+0.24}_{-0.51}$ & $-$14.88$^{+0.10}_{-0.14}$ & 36.68$^{+0.09}_{-0.12}$ & 36.47$^{+0.24}_{-0.51}$ & 36.27$^{+0.10}_{-0.14}$ & $-$0.39 & 1.73$^{+0.15}_{-0.14}$ \\
HCH99-18 & NGC 5128 & 843 & 3.1 & 0.13 & 3.9 & 3386.95$\pm$81.37 & 1279.31$\pm$57.69 & 2119.43$\pm$57.56 & $-$13.26$^{+0.02}_{-0.02}$ & $-$13.42$^{+0.03}_{-0.03}$ & $-$13.79$^{+0.02}_{-0.02}$ & 37.88$^{+0.02}_{-0.02}$ & 37.73$^{+0.03}_{-0.03}$ & 37.35$^{+0.02}_{-0.02}$ & $-$0.25 & 1.41$^{+0.04}_{-0.04}$ \\
HGHH92-C12=R281$^*$ & NGC 5128 & 216 & 8.2 & 1.43 & 16.2 & 69.48$\pm$42.24 & 25.31$\pm$36.06 & 44.04$\pm$22.08 & $-$14.50$^{+0.31}_{-0.79}$ & $-$14.38$^{+0.44}_{-0.95}$ & $-$15.01$^{+0.34}_{-0.86}$ & 36.64$^{+0.31}_{-0.79}$ & 36.76$^{+0.44}_{-0.95}$ & 36.13$^{+0.34}_{-0.86}$ & $-$0.27 & 1.49 \\
SUCD1 & Sombrero & 192 & 3.1 & 0.29 & 4.2 & 106.37$\pm$41.99 & 48.12$\pm$36.01 & 59.21$\pm$21.72 & $-$14.17$^{+0.25}_{-0.51}$ & $-$14.23$^{+0.33}_{-0.88}$ & $-$14.79$^{+0.23}_{-0.57}$ & 37.90$^{+0.25}_{-0.51}$ & 37.85$^{+0.33}_{-0.88}$ & 37.28$^{+0.23}_{-0.57}$ & $-$0.1 & 1.05$^{+0.23}_{-0.22}$ \\
M60-UCD1 & Virgo & 307 & 1.4 & 0.15 & 1.9 & 107.77$\pm$42.01 & 26.87$\pm$35.40 & 81.90$\pm$22.74 & $-$14.69$^{+0.21}_{-0.46}$ & $-$14.62$^{+0.42}_{-1.19}$ & $-$15.10$^{+0.20}_{-0.41}$ & 37.84$^{+0.21}_{-0.46}$ & 37.90$^{+0.42}_{-1.19}$ & 37.42$^{+0.20}_{-0.41}$ & $-$0.51 & 1.61$^{+0.20}_{-0.20}$ \\
M85-HCC1$^\dagger$ & Virgo & 39 & 0.7 & 0.15 & 3.1 & 22.68$\pm$40.46 & 0.54$\pm$35.05 & 22.14$\pm$20.30 & $-$14.25$^{+0.39}_{-0.95}$ & $-$14.02$^{+0.43}_{-1.00}$ & $-$14.68$^{+0.38}_{-0.94}$ & 38.28$^{+0.39}_{-0.95}$ & 38.51$^{+0.43}_{-1.00}$ & 37.85$^{+0.38}_{-0.94}$ & $-$0.95 & 1.49 \\
\end{tabular}}
\textbf{Notes:} Col (1): UCD name adopted from the literature. $*$ means that X-ray emission from the UCD was detected during the second tier search using individual rather than merged observations. $^\dagger$ indicates a borderline detection (net counts are consistent with zero); see \autoref{sec:borderline} for more information. Col (2): UCD host system. Col (3): Total exposure time [ksec]. Col (4): Exposure time-weighted average off-axis angle of all observations used for flux derivation [arcmin]. Col (5): Separation between the optical position of the UCD and the center of the cross-matched X-ray source's aperture as defined by \texttt{wavdetect} [arcsec]. Col (6): Exposure time-weighted average radius of the circular apertures used to enclose 90\% of the PSF in the individual observations [arcsec]. Col (7): Net counts in full band (0.5-7.0 keV) with $1\sigma$ \citet{gehrels86} error [counts]. Col (8): Net counts in hard band (2.0-7.0 keV) with $1\sigma$ \citet{gehrels86} error [counts]. Col (9): Net counts in soft band (0.5-2.0 keV) with $1\sigma$ \citet{gehrels86} error [counts]. Col (10): Flux in full band with $90\%$ confidence interval [erg s$^{-1}$ cm$^{-2}$]. Col (11): Flux in hard band with $90\%$ confidence interval [erg s$^{-1}$ cm$^{-2}$]. Col (12): Flux in soft band with $90\%$ confidence interval [erg s$^{-1}$ cm$^{-2}$]. Col (13): Luminosity in full band with $90\%$ confidence interval [erg s$^{-1}$]. Col (14): Luminosity in hard band with $90\%$ confidence interval [erg s$^{-1}$]. Col (15): Luminosity in soft band with $90\%$ confidence interval [erg s$^{-1}$]. Col (16): Classical hardness ratio: ($C_H-C_S$)/($C_H+C_S$). Col (17): Best-fit power law photon index (set to 1.49 -- the median of the 11 derived indices -- if $\gamma$ was fixed due to $NC_F<100$).
\end{sidewaystable*}

\section{Origin of the X-ray Emission}\label{sec:results}
The most immediate conclusion to be drawn from our archival study is that across thirteen different host systems that collectively host 578 UCDs, we find only 21 X-ray-emitting UCDs (all coincidentally spectroscopically-confirmed). Considering only the 336 spectroscopically-confirmed UCDs with usable archival X-ray data and imposing a completeness limit of $L_X>2\times10^{38}$ erg s$^{-1}$ (to account for the variable depths in the different host systems), we measure a global X-ray detection fraction of $4/149\approx3\%$. In this section, we explore links between the X-ray and physical properties of UCDs with the goal of determining the origin of the X-ray emission. 

\subsection{LMXBs versus Central Black Holes}\label{sec:lmxbintro}
Our main concern is to quantify the expected contribution from the two most relevant sources of X-ray emission in UCDs: LMXBs (see discussion in \autoref{sec:lmxb}), and weakly-accreting central black holes (see discussion in \autoref{sec:bhocc}). Here, we briefly discuss the expected properties of the X-ray emission if dominated by one or the other.

At luminosities below $\sim5\times10^{38}$ erg s$^{-1}$, the LMXB X-ray luminosity function of GCs steeply rises \citep[e.g.,][]{fabbiano06,kim09}). Thus, our X-ray detections at these low $L_X$ are likely dominated by accretion onto stellar-mass objects in LMXBs. At higher $L_X$, however, we can hope to be sensitive to accretion both onto central black holes and in LMXBs. For a given stellar mass, the LMXB fraction should rise with stellar density due to the increased probability of producing interacting stellar binaries \citep{jordan04}. In fact, as we will quantify in \autoref{sec:lmxb}, given the optical properties of the UCDs, we would expect many more LMXBs than our number of X-ray detections.

To estimate the likely emission from a central accreting black hole, consider that the Eddington luminosity\footnote{$L_{\rm edd}=1.3\times10^{38} (M_{\rm BH}/M_{\odot})$ erg s$^{-1}$.} of a $10^6M_{\odot}$ black hole is $\sim1.3\times10^{44}$ erg s$^{-1}$.\footnote{A $10^6M_{\odot}$ black hole would account for $\sim10\%$ of the typical UCD's dynamical mass, similar to the black hole mass fraction found in M60-UCD1 \citep{seth14}.} Adopting the relation from \citet{ho08} that $L_{\rm bol}\approx16\times L_{X}$ for the typical low-luminosity AGN, our typical X-ray upper limits of $10^{37}$ and $10^{39}$ erg s$^{-1}$ allow us to probe bolometric Eddington ratios ($\equiv L_{\rm bol}/L_{\rm edd}$) down to $\sim7\times10^{-8}$ and $\sim7\times10^{-6}$, respectively, assuming a $10^6M_{\odot}$ central black hole. Given that the X-ray-detected M60-UCD1's dynamically-confirmed SMBH is radiating at $L_{\rm bol}/L_{\rm edd}\approx4\times10^{-7}$, we certainly have the ability to detect some accreting central black holes if they exist.

The X-ray spectral and variability properties of UCDs can provide some additional clues about the nature of the X-ray emission. However, the median derived photon index (assuming an absorbed power law spectral model; $\gamma\approx1.5$) is consistent with both a low-luminosity AGN \citep{ho08} and an LMXB in the hard/low state \citep{fabbiano06,remillard06}. It is not clear what conditions would be required to produce the hard X-ray spectra ($\gamma\approx1$) of some UCDs (F-34, HGHH92-C21, and SUCD1). In terms of variability, the greatest argument in favor of LMXBs would come from short-term variability which would imply short crossing times. Alas, we find no evidence for short-term variability (on the order of ksec), and instead detect only long-term variability (spanning years). Since the data do not suggest any periodicity, the variability could arise from either a central black hole or a population of LMXBs.

The X-ray properties alone are therefore inconclusive in addressing the possible domination of LMXBs over central black holes. In contrast, considering both the X-ray and optical properties together may help reveal concrete differences between the LMXB and central black hole scenarios.

\subsection{X-ray Detection Fraction}\label{sec:xdet}
We now ask whether the X-ray detection fraction depends on the physical properties of the UCDs. If the most massive and compact UCDs with the highest stellar velocity dispersions hosted either central accreting black holes \citep[e.g.,][]{mieske13} or a large population of dark stellar remnants \citep[e.g.,][]{dabringhausen12}, then we would expect positive physical trends between the X-ray detection fraction and various physical properties. It is a bit complicated to address this vital question because we have such variable X-ray upper limits. Therefore, when computing the detection fraction at a given $L_X$ completeness limit, we include all X-ray-detected UCDs (above that completeness limit) as well as non-X-ray-detected UCDs whose upper limits are greater than the completeness limit. We bin the detections as a function of four different physical properties: absolute $V$-band magnitude, dynamical mass-to-light ratio in the $V$ band, global stellar velocity dispersion, and surface luminosity density ($\Sigma=\frac{L_V}{2\pi r_{\rm hl}^2}$). As a concrete lower bound on the detection fraction upper limits, we also show the actual detection fraction in each bin (without including the non-X-ray-detected UCDs), assuming a completeness limit of $L_X>10^{38}$ erg s$^{-1}$.

\begin{figure*}
\begin{center}
\includegraphics[width=1\hsize]{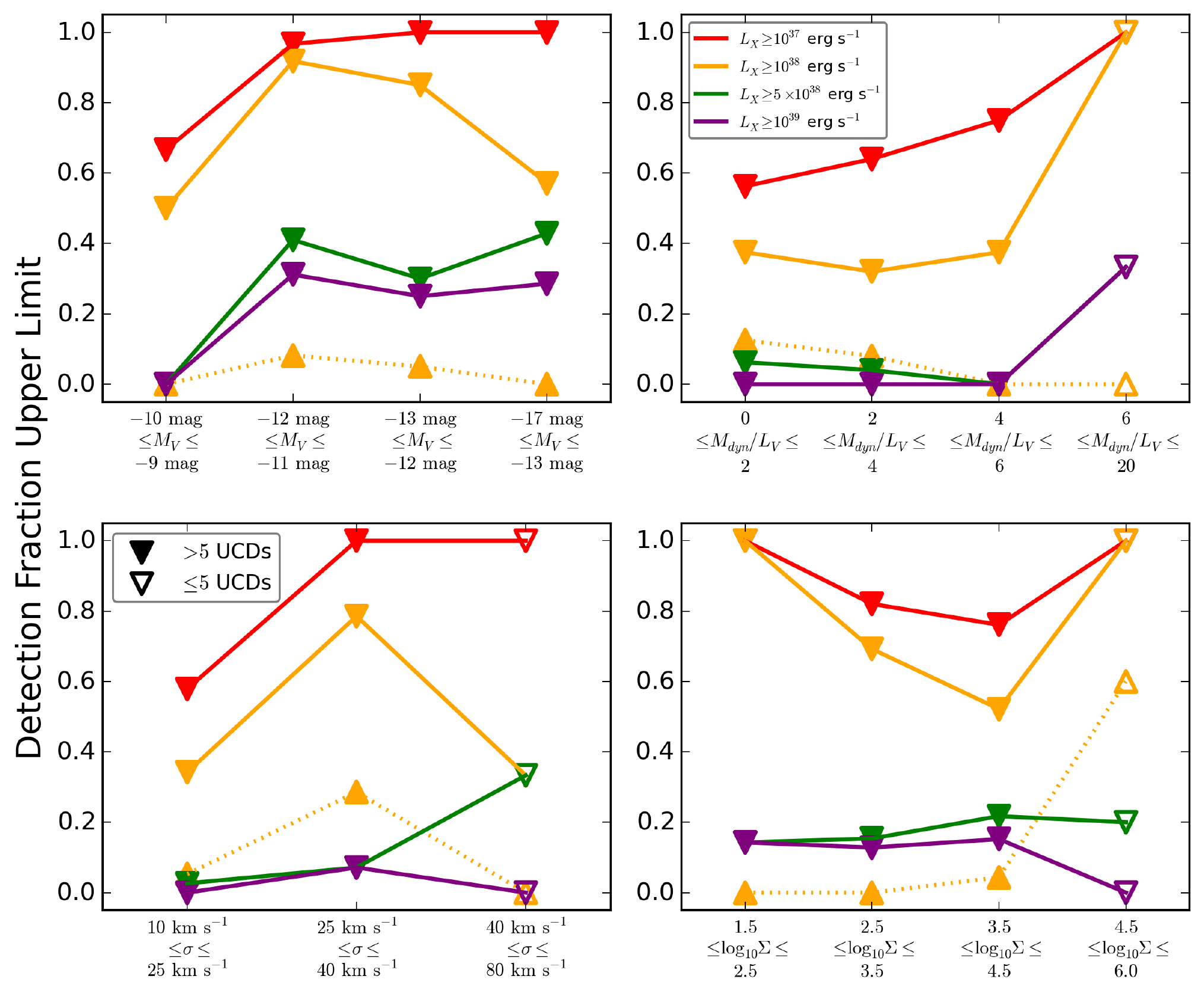}
\end{center}
\caption{Upper limits on the X-ray detection fraction as a function of four different physical properties using four different $L_X$ completeness limits. \textit{Clockwise from top-left:} Absolute $V$-band magnitude, dynamical mass-to-light ratio in the $V$-band, the surface luminosity density in the $V$-band, and the global stellar velocity dispersion. Filled triangles are used for bins with more than 5 UCDs, and empty triangles are used for bins with 5 or fewer UCDs. Also included is a ``lower bound" based on our actual detection fraction assuming a completeness limit of $L_X>10^{38}$ erg s$^{-1}$ (orange dotted lines and upward-facing triangles).\label{fig:lxfrac}}
\end{figure*}

As can be seen in \autoref{fig:lxfrac}, the upper limits on the detection fractions show an increasing trend between some bins for each measured property. However, the trends are not consistent for the different completeness limits that we explore, and they are also not consistent with the trends seen for the actual detection fractions assuming a completeness limit of $L_X>10^{38}$ erg s$^{-1}$.  More observations are needed, particularly at the high end of each physical property's distribution where the number of UCDs is lower. Tentatively, there does appear to be an increasing trend in the actual detection fractions (assuming a completeness limit of $L_X>10^{38}$ erg s$^{-1}$) toward higher bins of $\Sigma$ and $\sigma$, which is also seen in the detection fraction upper limits. There is also a hint of environmental dependence in our raw detection fractions, which peak for the isolated galaxy NGC 5128. However, this dependence is driven entirely by the non-uniform X-ray sensitivities arising from the different host system distances. In the future, it would be very interesting to determine whether the accretion rate is in fact dependent on environmental richness and/or proximity to a massive neighboring galaxy \citep[e.g.,][]{liu15b}. 

\subsection{X-ray--Physical Correlations and Statistics}\label{sec:xcorr}
Next, we search for: (1) correlations between $L_X$ and physical properties within our sample of 21 X-ray-detected UCDs, and (2) statistical differences in the distributions of physical properties between the X-ray-detected and non-X-ray-detected samples. The results of these two searches can reveal clues about the origin of the X-ray emission. For example, a positive correlation between $L_X$ and $M_V$ would suggest a direct connection to the stellar, rather than dark, component of UCDs, and thus lend credence to LMXBs as the dominant source of the X-ray emission. On the other hand, a positive correlation between $L_X$ and \mldyn\;would suggest a connection to the dark component \citep[of which central black holes are but one possibility;][]{mieske13}. Furthermore, if the \mldyn\;distribution of the X-ray-detected sample is offset toward higher values than that of the non-X-ray-detected sample, then that would also suggest a relationship between the X-ray emission and the dark mass in UCDs. 

\begin{figure*}
\begin{center}
\includegraphics[width=0.9\hsize]{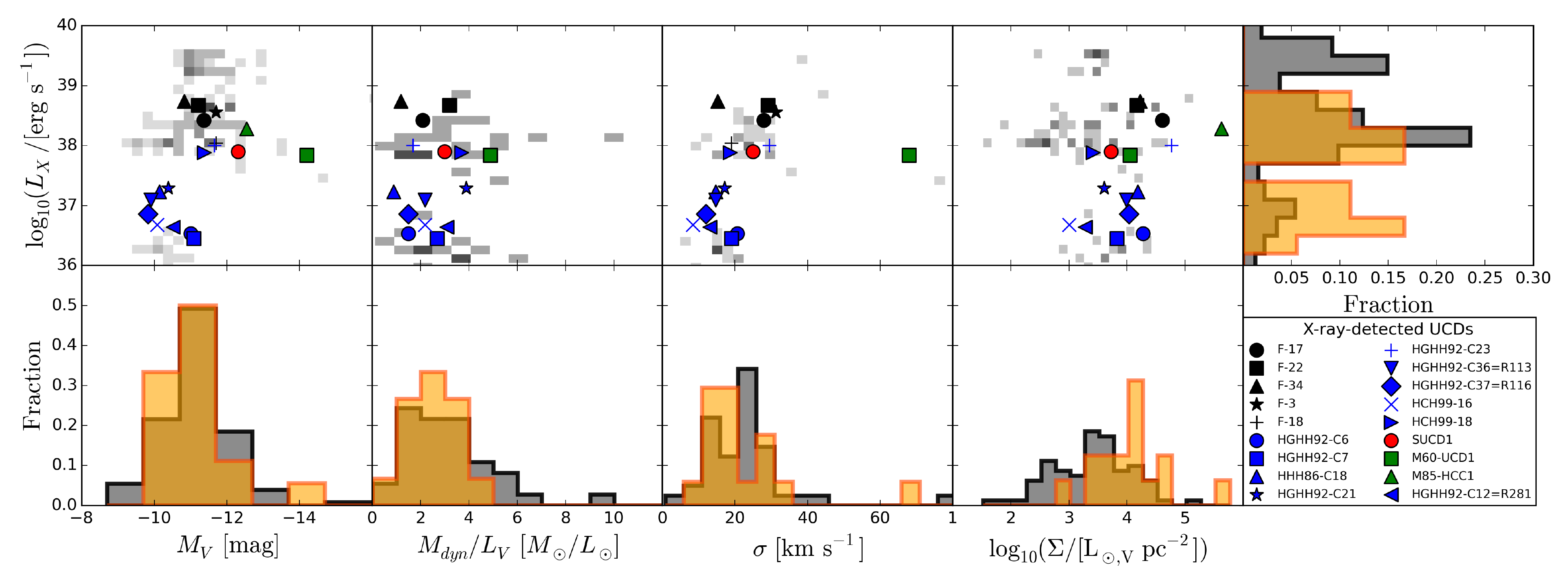}
\end{center}
\caption{The distributions of full-band (0.5-7.0 keV) $L_X$ and physical properties for our sample of 21 X-ray-detected UCDs. \textit{Top row:} The $L_X$ values plotted against four different physical properties: absolute $V$-band magnitude, dynamical mass-to-light ratio in the $V$-band, global stellar velocity dispersion, and surface luminosity density. The $L_X$ upper limits for the non-X-ray-detected UCDs are shown as a two-dimensional histogram in the background. The right-most subplot shows the distribution of $L_X$ itself: orange for X-ray-detected UCDs and gray for non-X-ray-detected UCDs. The UCDs in each host system are plotted with the same color: black (Fornax), blue (NGC 5128), red (Sombrero), and green (Virgo). \textit{Bottom row:} The distribution of physical properties for spectroscopically-confirmed X-ray-detected (orange) and non-X-ray-detected (gray) UCDs. The total number of UCDs in both samples with the required measurements available is given in \autoref{tab:nonpar}.\label{fig:lxdet}}
\end{figure*}

The top panels of \autoref{fig:lxdet} show the distribution of full-band (0.5-7.0 keV) $L_X$ as a function of various physical properties: $M_V$, \mldyn, $\sigma$, and surface luminosity density ($\Sigma$). The $L_X$ upper limits of non-X-ray-detected UCDs are shown via a two-dimensional histogram in the background. We see no obvious correlation between $L_X$ and \mldyn\;for the X-ray-detected UCDs. However, $L_X$ appears to be negatively correlated with $M_V$ and positively correlated with $\sigma$ (i.e., optically brighter and thus more massive X-ray-detected UCDs, with higher stellar velocity dispersions, tend to also have higher $L_X$). To quantify the strength of the linear correlation between each property and $L_{X,0.5-7.0}$, we computed the Pearson product-moment correlation coefficient (R) and its corresponding p-value, given in \autoref{tab:nonpar}. We excluded the outlier M60-UCD1 when computing R for $M_V$ and $\sigma$, and we excluded the outlier M85-HCC1 when computing R for $\Sigma$. The values of R indicate a moderate positive correlation with $\sigma$ at $99\%$ significance, and a moderate negative correlation with $M_V$ at $99\%$ significance. The positive correlation with $\Sigma$ is marginal (at $>90\%$ significance). There is no correlation with \mldyn\;($p=0.71$). 

In the bottom panels of \autoref{fig:lxdet}, we compare the same four physical properties for spectroscopically-confirmed X-ray-detected and non-X-ray-detected UCDs\footnote{We emphasize that most UCDs do not have these four physical properties available, so we must necessarily restrict comparisons to a small subset of the non-X-ray-detected UCDs. See \autoref{tab:nonpar}.}. From visual inspection alone, there are no systematic deviations between the two samples' distributions of $M_V$, \mldyn, and $\sigma$. However, there does appear to be a systematic offset between the surface luminosity density distributions: X-ray-detected UCDs tend to be more compact. We ran three different non-parametric tests to check for statistically-significant deviations between the two groups: the 2-sample Kolmogorov-Smirnov (KS) test, the Mann-Whitney-Wilcoxon (MWW) rank sums test, and the Anderson-Darling (AD) test. All three tests check whether two samples are consistent with being drawn from the same underlying parent population by making use of different test statistics. The number of X-ray-detected and non-X-ray-detected UCDs (with the required properties available) used for each test, and the resulting p-values, are given in \autoref{tab:nonpar}. Only for the surface luminosity density $\Sigma$ can we reject the null hypothesis that the X-ray-detected and non-X-ray-detected $\Sigma$ distributions are drawn from the same parent populations, at the $>99\%$ significance level ($p<0.01$), suggesting an LMXB origin for the X-rays \citep[consistent with][]{phillipps13}.

\begin{table}[!htbp]
\begin{center}
\caption{X-ray--Physical Statistics\label{tab:nonpar}}
\scriptsize
\begin{tabular}{ccccccc}
\tableline
Property & $N_X$ & $N_{non-X}$ & KS & MWW & AD & R (p) \\
\tableline
$M_V$ & 18 & 130 & 0.923 & 0.808 & 0.854 & -0.60 (0.01) \\
$M_{\mathrm{dyn}}/L_V$ & 15 & 37 & 0.292 & 0.207 & 0.287 & 0.11 (0.71) \\
$\sigma$ & 17 & 41 & 0.403 & 0.657 & 0.765 & 0.65 (0.01) \\ 
$L_V/(2\pi r_{\rm hl}^{2})$ & 16 & 81 & 0.004 & 0.001 & 0.001 & 0.47 (0.08) \\
\end{tabular}
\tablecomments{The p-values for each of the three non-parametric tests used to check whether the two samples of optical properties (X-ray-detected vs. non-X-ray-detected UCDs) are drawn from the same underlying parent population. The null hypothesis (that the two samples are drawn from the same underlying parent population) can only be rejected in the case of compactness, which shows statistical evidence for a discrepancy at the $>99\%$ level. The Pearson product-moment correlation coefficient (R) and its corresponding p-value are also given to quantify the strength of each property's correlation with $L_{X,0.5-7.0}$.}
\end{center}
\end{table}

\subsection{Contribution from LMXBs}\label{sec:lmxb}
We now calculate the expected X-ray detection fraction from LMXBs alone. Note that HMXBs, which can have the same X-ray luminosities as LMXBs but which comprise O- and B-type stars, are not expected to contribute to the X-ray emission from UCDs' old stellar systems. Furthermore, since all of our X-ray-detected UCDs have peak $L_X<10^{39}$ erg s$^{-1}$, they are by definition too faint to be ULXs \citep{fabbiano06}. Our best means to estimate the LMXB rate in UCDs is to scale from GCs, for which far more is known. In GCs, as the stellar density increases, the LMXB rate per unit stellar mass also increases because of the higher probability of forming interacting stellar binaries \citep{jordan04}. The stellar encounter rate, $\Gamma$, scales as $M_*^{1.5} r_{\rm hl}^{-2.5}$ \citep[see derivation in][]{dabringhausen12}. This relation is generally invoked to explain why the vast majority of LMXBs in the (local) Universe are found in GCs and only a small fraction are in dwarf and massive galaxies \citep[e.g.,][]{fabbiano06}. 

It is not yet known how UCDs fit into this picture since UCDs are far more compact than typical dwarf and massive galaxies, but less compact than GCs. \citet{phillipps13} noted that if the GC scaling relation were to be applied to UCDs, then the $\sim10\times$ larger mass and size of UCDs would, in general, tend to cancel, giving rise to a similar LMXB detection fraction for UCDs as for GCs. The LMXB detection fraction among the entire GC population (without splitting into red (metal-rich) and blue (metal-poor) GC subpopulations) is $\sim6.9\%$ \citep[down to $L_X\approx2\times10^{37}$ erg s$^{-1}$;][]{kim13}. In contrast, the LMXB occurrence rate in non-nucleated massive galaxies is far lower than in nucleated dwarf galaxies and GCs, and the scaling relations for non-nucleated massive galaxies are derived with stellar mass rather than stellar density \citep[e.g.,][]{gilfanov04,kim06}. At the moment, scaling to GCs is the best method that we have for obtaining a fiducial benchmark for UCDs, but the analogy is imperfect given their different star formation histories and stellar densities. 

To quantify the occurrence of LMXBs in our sample of UCDs, we employ equation (17) of \citet{sivakoff07} which gives the expected number of LMXBs: 

\begin{equation}
\scriptsize n_{\rm LMXB} = 8.0\times10^{-2} \left(\frac{M}{10^6M_{\odot}}\right)^{1.237} 10^{0.90(g-z)}\left(\frac{r_{\rm hl}}{1\;\mathrm{pc}}\right)^{-2.22}\;.
\end{equation}
Again, this scaling is not perfectly matched to our needs. Not only did \citet{sivakoff07} derive their equation (17) using GCs rather than UCDs, but they also used a luminosity-limited sample with a completeness limit of $L_X>3.2\times10^{38}$ erg s$^{-1}$. Although we do not know $M_{\rm star}/L_V$ and $(g-z)$ for each individual UCD, we make the following educated guesses. We convert $L_V$ to stellar mass assuming $M_{\rm star}/L_V=4$ because UCDs in the literature have stellar mass-to-light ratios between 3 and 5. We assume $(g-z)=1$ for all the UCDs based on the finding by \citet{zhang15} that UCDs around M87 in Virgo are tightly clustered around that value. Including the color/metallicity is important because blue (metal-poor) GCs are less likely to host an LMXB than are red (metal-rich) GCs \citep[e.g.,][]{kim13}. Of the 16 X-ray-detected and 81 non-X-ray-detected UCDs with $M_V$ and $r_{\rm hl}$ available, 21 non-X-ray-detected UCDs have upper limits greater than the \citet{sivakoff07} completeness limit, suggesting that we would not be sensitive to LMXBs with $L_X\approx3.2\times10^{38}$ erg s$^{-1}$ in those 21 UCDs. Thus, we calculate $n_{\rm LMXB}$ for the remaining 76 UCDs, and show $n_{\rm LMXB}$ for the 16 of these that are X-ray-detected in \autoref{tab:optdet}. 

Taking a statistical approach \citep[e.g.,][]{gilfanov04stat}, when we sum the $n_{\rm LMXB}$ estimates for the 16 X-ray-detected UCDs as well as the 60 non-X-ray-detected UCDs with upper limits $<3.2\times10^{38}$ erg s$^{-1}$, we expect $\sim25$ LMXBs with $L_X>3.2\times10^{38}$ erg s$^{-1}$. However, we have detected only 2 UCDs with $L_{X}>3.2\times10^{38}$ erg s$^{-1}$. It is difficult to uniquely interpret this deficiency of detections (an excess of detections might have hinted at central black holes), but the most conservative interpretation is that LMXBs are responsible for the X-ray emission. The discrepancy between the expected $34\%$ LMXB fraction and the actual $3\%$ detection fraction reveals a difference between the bright ($L_X>3.2\times10^{38}$ erg s$^{-1}$) ends of the GC-LMXB and UCD-LMXB X-ray luminosity functions (see \autoref{sec:deficiency}).

To sum up, our order of magnitude estimate for the expected number of LMXBs already suggests that all of the observed X-rays are consistent with coming from LMXBs. This conclusion is bolstered by the rising X-ray detection fraction with increasing surface luminosity density (\autoref{fig:lxfrac}), and the tendency of X-ray-detected UCDs to be more compact than non-X-ray-detected UCDs (\autoref{fig:lxdet}). The X-ray spectra and long-term time variability are also consistent with LMXBs, as discussed in \autoref{sec:lmxbintro}. Taken together, our sample of X-ray-detected UCDs is compatible with a scenario in which LMXBs produce, and indeed dominate, the observed X-ray emission. The study of \citet{phillipps13}, based on Fornax alone, came to similar conclusions.

\section{Discussion}\label{sec:disc}

\subsection{Deficiency of LMXBs in UCDs}\label{sec:deficiency}
In \autoref{sec:lmxb}, we showed that the number of X-ray-detected UCDs is less than the expected number of LMXBs down to the $L_X$ completeness limit of \citet{sivakoff07}. Whereas we expected to detect $\sim25$ UCDs with $L_X>3.2\times10^{38}$ erg s$^{-1}$, we detected only two UCDs above that X-ray luminosity. The derivation of $\sim25$ expected LMXBs was predicated upon the assumption that UCDs follow the same scaling relation as GCs between the number of LMXBs, the stellar density $\Gamma$, and other intrinsic properties such as color/metallicity. The deficiency of X-ray detections compared to the \citet{sivakoff07} prediction may be a clue that UCDs are not as efficient as GCs in producing LMXBs, even after controlling for stellar density. This lower LMXB production efficiency in UCDs may be the result of different metallicities, mean stellar ages, and/or structural properties as compared to GCs. In the future, it will be interesting to investigate whether this discrepancy holds at $L_X<3.2\times10^{38}$ erg s$^{-1}$. 

\subsection{UCD Central BH Occupation Fraction}\label{sec:bhocc}
The original aim of this paper was to provide X-ray constraints on the central black hole occupation fraction of UCDs. However, any such X-ray constraints would be weak because, as an ensemble, UCDs are thought to host a large number of LMXBs whose collective X-ray emission may dominate over weakly-accreting central black holes, as shown in \autoref{sec:lmxb}. Unfortunately, even if a massive central black hole was present in a UCD, it may not necessarily be accreting, or it may be accreting below our detection limits.

We can address the detectability of weakly-accreting central black holes as follows. Suppose that all UCDs contain a central black hole with $M_{\rm BH}/M_{\rm star}=10\%$, where $M_{\rm star}$ is derived from $M_V$ as in \autoref{sec:lmxb}. By adopting such a high BH mass fraction \citep[compared to the $\sim0.1\%$ seen in massive galaxies; e.g.,][]{haringrix04}, we will explore the region of parameter space that may be inhabited by extreme UCDs which are like M60-UCD1 and which may also host ``overweight" central black holes \citep{seth14,sandoval15}. Our adopted BH mass fraction is also tantamount to assuming that all UCDs are tidally-stripped nuclei whose progenitors had $M_{\rm star}\sim10^{10}M_{\odot}$. Of course, the dynamical modeling of \citet{mieske13} already suggests that only a small subset of UCDs host central black holes with $M_{\rm BH}/M_{\rm star}\gg0.1\%$ \citep[see also][]{seth14}. Nevertheless, estimated in this way, the median assumed $M_{\rm BH}\approx10^6M_{\odot}$ and the median assumed limiting $L_X/L_{\rm edd}\approx5\times10^{-7}$. 

A significant obstacle in our calculations is that we do not know a priori the true distribution of accretion rates, which we need in order to predict the X-ray luminosities corresponding to our assumed distribution of BH masses. Our best guess, perhaps, comes from the measurements of \citet{miller12}, who find that $\sim50\%$ of bulge-dominated galaxies with $M_{\rm star}\sim10^{10}M_{\odot}$ are radiating at $L_X/L_{\rm edd}\geq10^{-7}$. Since we are assuming that all UCDs are the tidally-stripped nuclei of more massive galaxies, we can use the population studied by \citet{miller12} as a proxy and consider the detectability of central black holes accreting at $L_X/L_{\rm edd}\geq10^{-7}$. From our study, we would be sensitive to such weakly-accreting central black holes in 29 UCDs, all of which have an assumed limiting $L_X/L_{\rm edd}\leq10^{-7}$. Presupposing that $50\%$ of those 29 hypothetical central black holes are actually radiating at $L_{X}/L_{\rm edd}\geq10^{-7}$ \citep[based on][]{miller12}, we would expect to detect X-ray emission from roughly 14 central black holes. In fact, among the original 29 UCDs are 12 X-ray-detected UCDs, which is not so different from the expected number of 14 X-ray-detectable central black holes. However, since these 12 X-ray-detected UCDs all have $L_X\leq10^{38}$ erg s$^{-1}$, we are again pushing into the faint LMXB-dominated X-ray regime and thus require complementary constraints from other types of measurements (e.g., optical emission-line spectroscopy and radio imaging).

\subsection{Complementary Radio Constraints}\label{sec:sucd1}
Although LMXBs and central black holes are not mutually exclusive, we are hitting a limit in the effectiveness of X-ray observations to distinguish between the two scenarios. One hopeful way forward is to look at the ratio of radio to X-ray luminosity, which is a strongly increasing function of $M_{\rm BH}$. This so-called fundamental plane of black hole activity \citep[e.g.,][]{merloni03,plotkin12} has been used to search for intermediate-mass black holes in GCs \citep[e.g.,][]{maccarone05,strader12b,strader12,chomiuk13} and SMBHs in other local galaxies \citep[e.g.,][]{reines11,reines14}. It is beyond the scope of this paper to search for archival radio observations of all non-X-ray-detected UCDs. However, for the 21 X-ray-detected UCDs, we did search the VLA Data Archive\footnote{\url{https://archive.nrao.edu/archive/advquery.jsp}} for serendipitous \textit{Jansky Very Large Array} (\textit{JVLA}) imaging at 1.4 GHz $<\nu<$ 10 GHz. Our search was restricted to the A- and B-array configurations, which provide the highest spatial resolution at 1.4 GHz $<\nu<$ 10 GHz. The high spatial resolution is important for distinguishing UCDs, which typically have sub-arcsec optical half-light radii, from nearby sources. Only three of the 21 X-ray-detected UCDs have archival \textit{JVLA} imaging, and only one of those is public: observations of the Sombrero galaxy (Proposal 12B-007; PI: Zhiyuan Li). 

\begin{figure*}
\begin{center}
\includegraphics[width=0.9\hsize]{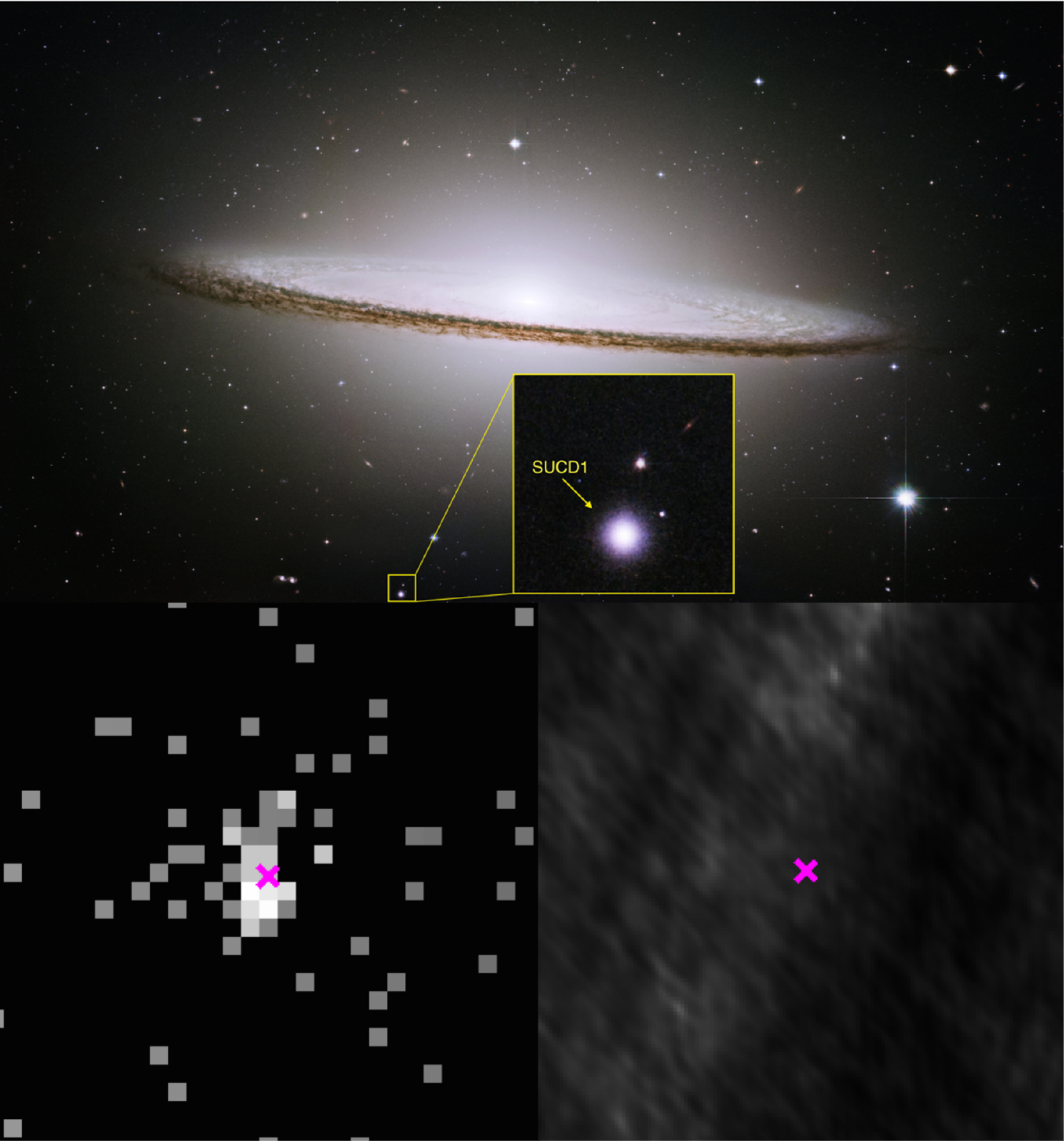}
\end{center}
\caption{The multi-wavelength case study of SUCD1 near the Sombrero galaxy. \textit{(Top)}: \textit{HST/ACS} (\textit{BVR}) image of Sombrero with SUCD1 identified via the $15"\times15"$ inset. \textit{(Bottom left)}: Chandra hard band (2.0-7.0 keV) image ($15"\times15"$) of SUCD1 showing a clear hard X-ray detection. \textit{(Bottom right)}: JVLA A-configuration C-band (4-6 GHz) image ($15"\times15"$) at location of SUCD1 showing a $\sim5$ GHz non-detection. In both the Chandra and JVLA images, the magenta `X' marks the optical position of SUCD1. \textit{HST} image from NASA and The Hubble Heritage Team (STScI/AURA).\label{fig:sucd1}}
\end{figure*}

SUCD1 is 2.84 arcmin from the primary beam center in a 2.5 hour integration at 5 GHz (C-band; single passband of width 2 GHz). \citet{hau09}, who discovered SUCD1, noted the presence of off-center X-ray emission from archival \textit{Chandra} observations, and claimed that the most likely origin of the X-ray emission is from LMXBs. Our study adds 176.2 ksec of new \textit{Chandra} observations taken in 2008 and not considered in \citet{hau09}. Although the additional \textit{Chandra} imaging allows us to more finely constrain the X-ray properties of SUCD1, the archival \textit{JVLA} imaging was taken $\sim4$ years after the \textit{Chandra} data, and so the two sets of observations may not track the same states of an accreting black hole, if one is present \citep[e.g.,][]{remillard06}. In \autoref{fig:sucd1}, we show the \textit{HST} (optical), \textit{Chandra} (hard X-ray), and \textit{JVLA} (5 GHz) images of SUCD1. There is no 5 GHz signal at the optical position of SUCD1 whereas there is clearly hard X-ray emission. We measured the 5 GHz RMS within a circular aperture of radius 1 arcsec centered at the optical position of SUCD1 and found it to be 6.77 $\mu$Jy. We adopted twice this RMS value as our upper limit on the 5 GHz flux density from SUCD1, and computed the corresponding upper limit on the 5 GHz radio luminosity: $L_{R,5}\lesssim 8.0\times10^{33}$ erg s$^{-1}$. 

With our measured $L_{X,2-7}$ and upper limit on $L_{R,5}$ in hand, we can use the fundamental plane of black hole activity to estimate an upper limit for the mass of a central accreting black hole in SUCD1, assuming that one was present and that all of the hard X-ray emission is produced by that hypothetical object.\footnote{From \citet{merloni03}: $\log_{10} L_R = 0.60 \log_{10} L_X + 0.78 \log_{10} M_{\rm BH} + 7.33$, where $L_R$ and $L_X$ have units of erg s$^{-1}$.} We arrive at an upper limit of $M_{\rm BH}\lesssim10^5M_{\odot}$, which would correspond to $\sim0.5\%$ of the dynamical mass of SUCD1 and yield $L_{X,0.5-7.0}/L_{\rm edd}\approx6\times10^{-6}$, assuming that all of the X-ray emission is produced by a central black hole.

Building on the constraint that we were able to place on SUCD1's hypothetical central black hole, we now examine the potential of a large-scale follow-up radio imaging campaign. Since X-ray constraints suggest that most UCDs have $M_{\rm BH}/M_{\rm star}<10\%$, as discussed in \autoref{sec:bhocc}, let us suppose that, rather than LMXBs, the 21 X-ray-detected UCDs host central black holes with $M_{\rm BH}/M_{\rm star}=0.1\%$. Assuming the fundamental plane of black hole activity, as in \autoref{sec:sucd1}, the predicted 5 GHz flux densities for X-ray-detected UCDs range from $\sim1\mu \rm Jy$ to $\sim37 \mu \rm Jy$, with a median of $\sim3 \mu \rm Jy$. A similar calculation for the non-X-ray-detected UCDs, using their $L_X$ upper limits, yields predicted 5 GHz flux density upper limits ranging from $\sim0.2\mu \rm Jy$ to $\sim8\mu \rm Jy$, with a median of $\sim1\mu \rm Jy$. RMS levels of $\sim2\mu \rm Jy$ and $\sim1\mu \rm Jy$ can be achieved in $\sim3$ hours and $\sim11$ hours, respectively, with the \textit{JVLA} in the A-configuration at 5 GHz (contrast this with the hundreds of hours worth of \textit{Chandra} exposure time needed to reach the $10^{37}$ erg s$^{-1}$ sensitivities presented in this paper). There are a handful of reasonably detectable UCDs assuming a $2\mu \rm Jy$ RMS level and a $3\sigma$ detection threshold. Several more radio-faint ($<6\mu \rm Jy$) UCDs become potentially detectable for $M_{\rm BH}/M_{\rm star}\approx1\%$. Among them are M60-UCD1, M59-UCD3 and M59cO, the three most extreme UCDs in Virgo that are the targets of the other two existing, but proprietary, \textit{JVLA} datasets (Proposal 15A-091, PI: Strader; Proposal 15A-154, PI: Mao). We have thus initiated a 5 GHz pilot radio survey of X-ray-detected UCDs in Fornax and NGC 5128 using the \textit{JVLA} (Proposal 16A-382, PI: Pandya). 

Although surveying individual UCDs may be prohibitive due to their faint 5 GHz emission, it would be interesting to consider a deep radio survey of the center of a nearby galaxy cluster (e.g., Virgo or Fornax). Then, an $\sim11$ hour exposure could have many science goals, including a survey of radio emission from UCDs. Given that the primary beam width at 5 GHz is $\sim9$ arcmin, targeting the center of Fornax (NGC 1399) would simultaneously observe 35/78 ($\approx45\%$) Fornax UCDs, of which 6 are X-ray-detected. Targeting the center of Virgo (M87) would simultaneously observe 45/114 ($\approx39\%$) Virgo UCDs, but there are no X-ray-detected UCDs around M87. In any case, by combining X-rays with deep radio imaging and extending our work on the prototypical SUCD1, we can hope to place a powerful constraint on the central (active) black hole occupation fraction of UCDs.

\section{Summary}\label{sec:conc}
In this paper, we have presented a comprehensive archival study of the X-ray properties of 578 UCDs distributed throughout thirteen different host systems, with the goal of finding weakly-accreting central black holes. Although we found 21 X-ray-detected UCDs with $L_{X,0.5-7.0}$ ranging from $2.8\times10^{36}$ erg s$^{-1}$ to $5.5\times10^{38}$ erg s$^{-1}$, we showed that their X-ray emission is fully consistent with an origin in LMXBs. Interestingly, however, the number of X-ray detections in our ensemble of UCDs is far less than the expected number of LMXBs, and this deficiency of LMXBs in UCDs hints at a possible discrepancy between the formation histories of UCDs and GCs. Given that LMXBs and weakly-accreting central black holes are not necessarily mutually exclusive, we addressed the potential of a complementary radio study to augment the results of our archival X-ray study. 

Our main conclusions are:
\begin{itemize}
\item Of the 336 spectroscopically-confirmed UCDs, 21 are X-ray-detected; imposing a completeness limit of $L_X>2\times10^{38}$ erg s$^{-1}$ to account for the variable depths of the different host systems, the global X-ray detection fraction is $4/149\approx3\%$. 
\item Seven of the 21 X-ray-detected UCDs show evidence of long-term X-ray time variability on the order of months to years.
\item The X-ray emission from X-ray-detected UCDs is consistent with arising from a population of LMXBs. 
\item The number of X-ray detections, assuming an LMXB origin, is lower than we would expect based on the UCDs' stellar densities (using a naive extrapolation from GCs). 
\item SUCD1 near the Sombrero galaxy is detected at 2-7 keV with \textit{Chandra} but not detected at 5 GHz with JVLA; assuming the fundamental plane of black hole activity, we place an upper limit on the mass of its hypothetical central black hole of $\lesssim10^5M_{\odot}$. 
\item A complementary archival or follow-up radio observing campaign may help identify weakly-accreting central black holes in UCDs via the fundamental plane of black hole activity.
\end{itemize}

\section*{Acknowledgements}
We thank Zhiyuan Li for providing a fully-reduced \textit{JVLA} image of Sombrero, Steffen Mieske for sending us the optical positions of UCDs in NGC 1023, and Andy Goulding for his suggestion to look at 4-7 keV images in Perseus. We are also grateful to the anonymous referee, the anonymous statistics expert, Michael Strauss, and Steffen Mieske for their thorough comments that improved the quality of this paper. In addition, VP greatly thanks: (1) Ken Glotfelty and Nicholas Lee at the \textit{CXC Help Desk} for answering his questions about \textit{Chandra} data manipulation, (2) Tony Perreault and Emmanuel Momjian at the NRAO Help Desk for their guidance on retrieving and analyzing archival \textit{JVLA} data, and (3) Michael Strauss, in whose AST 303 course VP originally wrote a final paper describing the initial motivation for carrying out this work.


\begin{thebibliography}{}
\expandafter\ifx\csname natexlab\endcsname\relax\def\natexlab#1{#1}\fi

\bibitem[{{Bergond} {et~al.}(2007){Bergond}, {Athanassoula}, {Leon},
  {Balkowski}, {Cayatte}, {Chemin}, {Guzm{\'a}n}, {Meylan}, \&
  {Prugniel}}]{bergond07}
{Bergond}, G., {Athanassoula}, E., {Leon}, S., {et~al.} 2007, \aap, 464, L21

\bibitem[{{Brodie} {et~al.}(2011){Brodie}, {Romanowsky}, {Strader}, \&
  {Forbes}}]{brodie11}
{Brodie}, J.~P., {Romanowsky}, A.~J., {Strader}, J., \& {Forbes}, D.~A. 2011,
  \aj, 142, 199

\bibitem[{{Caso} {et~al.}(2013){Caso}, {Bassino}, {Richtler}, {Smith Castelli},
  \& {Faifer}}]{caso13}
{Caso}, J.~P., {Bassino}, L.~P., {Richtler}, T., {Smith Castelli}, A.~V., \&
  {Faifer}, F.~R. 2013, \mnras, 430, 1088

\bibitem[{{Chiboucas} {et~al.}(2011){Chiboucas}, {Tully}, {Marzke},
  {Phillipps}, {Price}, {Peng}, {Trentham}, {Carter}, \&
  {Hammer}}]{chiboucas11}
{Chiboucas}, K., {Tully}, R.~B., {Marzke}, R.~O., {et~al.} 2011, \apj, 737, 86

\bibitem[{{Chomiuk} {et~al.}(2013){Chomiuk}, {Strader}, {Maccarone},
  {Miller-Jones}, {Heinke}, {Noyola}, {Seth}, \& {Ransom}}]{chomiuk13}
{Chomiuk}, L., {Strader}, J., {Maccarone}, T.~J., {et~al.} 2013, \apj, 777, 69

\bibitem[{{Da Rocha} {et~al.}(2011){Da Rocha}, {Mieske}, {Georgiev}, {Hilker},
  {Ziegler}, \& {Mendes de Oliveira}}]{darocha11}
{Da Rocha}, C., {Mieske}, S., {Georgiev}, I.~Y., {et~al.} 2011, \aap, 525, A86

\bibitem[{{Dabringhausen} {et~al.}(2012){Dabringhausen}, {Kroupa},
  {Pflamm-Altenburg}, \& {Mieske}}]{dabringhausen12}
{Dabringhausen}, J., {Kroupa}, P., {Pflamm-Altenburg}, J., \& {Mieske}, S.
  2012, \apj, 747, 72

\bibitem[{{Drinkwater} {et~al.}(2003){Drinkwater}, {Gregg}, {Hilker}, {Bekki},
  {Couch}, {Ferguson}, {Jones}, \& {Phillipps}}]{drinkwater03}
{Drinkwater}, M.~J., {Gregg}, M.~D., {Hilker}, M., {et~al.} 2003, \nat, 423,
  519

\bibitem[{{Drinkwater} {et~al.}(2000){Drinkwater}, {Jones}, {Gregg}, \&
  {Phillipps}}]{drinkwater00}
{Drinkwater}, M.~J., {Jones}, J.~B., {Gregg}, M.~D., \& {Phillipps}, S. 2000,
  \pasa, 17, 227

\bibitem[{{Evans} {et~al.}(2010){Evans}, {Primini}, {Glotfelty}, {Anderson},
  {Bonaventura}, {Chen}, \& {Davis}}]{evans10}
{Evans}, I.~N., {Primini}, F.~A., {Glotfelty}, K.~J., {et~al.} 2010, \apjs,
  189, 37

\bibitem[{{Evstigneeva} {et~al.}(2007){Evstigneeva}, {Gregg}, {Drinkwater}, \&
  {Hilker}}]{evstigneeva07}
{Evstigneeva}, E.~A., {Gregg}, M.~D., {Drinkwater}, M.~J., \& {Hilker}, M.
  2007, \aj, 133, 1722

\bibitem[{{Evstigneeva} {et~al.}(2008){Evstigneeva}, {Drinkwater}, {Peng},
  {Hilker}, {De Propris}, {Jones}, {Phillipps}, {Gregg}, \&
  {Karick}}]{evstigneeva08}
{Evstigneeva}, E.~A., {Drinkwater}, M.~J., {Peng}, C.~Y., {et~al.} 2008, \aj,
  136, 461

\bibitem[{{Fabbiano}(2006)}]{fabbiano06}
{Fabbiano}, G. 2006, \araa, 44, 323

\bibitem[{{Fellhauer} \& {Kroupa}(2005)}]{fellhauer05}
{Fellhauer}, M., \& {Kroupa}, P. 2005, \mnras, 359, 223

\bibitem[{{Firth} {et~al.}(2008){Firth}, {Drinkwater}, \& {Karick}}]{firth08}
{Firth}, P., {Drinkwater}, M.~J., \& {Karick}, A.~M. 2008, \mnras, 389, 1539

\bibitem[{{Forbes} {et~al.}(2013){Forbes}, {Pota}, {Usher}, {Strader},
  {Romanowsky}, {Brodie}, {Arnold}, \& {Spitler}}]{forbes13}
{Forbes}, D.~A., {Pota}, V., {Usher}, C., {et~al.} 2013, \mnras, 435, L6

\bibitem[{{Francis} {et~al.}(2012){Francis}, {Drinkwater}, {Chilingarian},
  {Bolt}, \& {Firth}}]{francis12}
{Francis}, K.~J., {Drinkwater}, M.~J., {Chilingarian}, I.~V., {Bolt}, A.~M., \&
  {Firth}, P. 2012, \mnras, 425, 325

\bibitem[{{Frank} {et~al.}(2011){Frank}, {Hilker}, {Mieske}, {Baumgardt},
  {Grebel}, \& {Infante}}]{frank11}
{Frank}, M.~J., {Hilker}, M., {Mieske}, S., {et~al.} 2011, \mnras, 414, L70

\bibitem[{{Freeman} {et~al.}(2001){Freeman}, {Doe}, \&
  {Siemiginowska}}]{sherpa01}
{Freeman}, P., {Doe}, S., \& {Siemiginowska}, A. 2001, in Society of
  Photo-Optical Instrumentation Engineers (SPIE) Conference Series, Vol. 4477,
  Astronomical Data Analysis, ed. J.-L. {Starck} \& F.~D. {Murtagh}, 76--87

\bibitem[{{Fruscione} {et~al.}(2006){Fruscione}, {McDowell}, {Allen},
  {Brickhouse}, {Burke}, {Davis}, {Durham}, {Elvis}, {Galle}, \&
  {Harris}}]{fruscione06}
{Fruscione}, A., {McDowell}, J.~C., {Allen}, G.~E., {et~al.} 2006, in Society
  of Photo-Optical Instrumentation Engineers (SPIE) Conference Series, Vol.
  6270, Society of Photo-Optical Instrumentation Engineers (SPIE) Conference
  Series, 1

\bibitem[{{Gallo} {et~al.}(2010){Gallo}, {Treu}, {Marshall}, {Woo}, {Leipski},
  \& {Antonucci}}]{gallo10}
{Gallo}, E., {Treu}, T., {Marshall}, P.~J., {et~al.} 2010, \apj, 714, 25

\bibitem[{{Gehrels}(1986)}]{gehrels86}
{Gehrels}, N. 1986, \apj, 303, 336

\bibitem[{{Gilfanov}(2004)}]{gilfanov04}
{Gilfanov}, M. 2004, \mnras, 349, 146

\bibitem[{{Gilfanov} {et~al.}(2004){Gilfanov}, {Grimm}, \&
  {Sunyaev}}]{gilfanov04stat}
{Gilfanov}, M., {Grimm}, H.-J., \& {Sunyaev}, R. 2004, \mnras, 351, 1365

\bibitem[{{Gregg} {et~al.}(2009){Gregg}, {Drinkwater}, {Evstigneeva}, {Jurek},
  {Karick}, {Phillipps}, {Bridges}, {Jones}, {Bekki}, \& {Couch}}]{gregg09}
{Gregg}, M.~D., {Drinkwater}, M.~J., {Evstigneeva}, E., {et~al.} 2009, \aj,
  137, 498

\bibitem[{{Gregory} \& {Loredo}(1992)}]{glvary92}
{Gregory}, P.~C., \& {Loredo}, T.~J. 1992, \apj, 398, 146

\bibitem[{{Ha{\c s}egan} {et~al.}(2005){Ha{\c s}egan}, {Jord{\'a}n},
  {C{\^o}t{\'e}}, {Djorgovski}, {McLaughlin}, {Blakeslee}, {Mei}, {West},
  {Peng}, {Ferrarese}, {Milosavljevi{\'c}}, {Tonry}, \& {Merritt}}]{hasegan05}
{Ha{\c s}egan}, M., {Jord{\'a}n}, A., {C{\^o}t{\'e}}, P., {et~al.} 2005, \apj,
  627, 203

\bibitem[{{H{\"a}ring} \& {Rix}(2004)}]{haringrix04}
{H{\"a}ring}, N., \& {Rix}, H.-W. 2004, \apjl, 604, L89

\bibitem[{{Hau} {et~al.}(2009){Hau}, {Spitler}, {Forbes}, {Proctor}, {Strader},
  {Mendel}, {Brodie}, \& {Harris}}]{hau09}
{Hau}, G.~K.~T., {Spitler}, L.~R., {Forbes}, D.~A., {et~al.} 2009, \mnras, 394,
  L97

\bibitem[{{Hilker}(2011)}]{hilker11}
{Hilker}, M. 2011, in EAS Publications Series, Vol.~48, EAS Publications
  Series, ed. M.~{Koleva}, P.~{Prugniel}, \& I.~{Vauglin}, 219--224

\bibitem[{{Hilker} {et~al.}(2007){Hilker}, {Baumgardt}, {Infante},
  {Drinkwater}, {Evstigneeva}, \& {Gregg}}]{hilker07}
{Hilker}, M., {Baumgardt}, H., {Infante}, L., {et~al.} 2007, \aap, 463, 119

\bibitem[{{Hilker} {et~al.}(1999){Hilker}, {Infante}, \& {Richtler}}]{hilker99}
{Hilker}, M., {Infante}, L., \& {Richtler}, T. 1999, \aaps, 138, 55

\bibitem[{{Ho}(2008)}]{ho08}
{Ho}, L.~C. 2008, \araa, 46, 475

\bibitem[{{Janz} {et~al.}(2015){Janz}, {Forbes}, {Norris}, {Strader}, {Penny},
  {Fagioli}, \& {Romanowsky}}]{janz15}
{Janz}, J., {Forbes}, D.~A., {Norris}, M.~A., {et~al.} 2015, \mnras, 449, 1716

\bibitem[{{Jennings} {et~al.}(2014){Jennings}, {Strader}, {Romanowsky},
  {Brodie}, {Arnold}, {Lin}, {Irwin}, {Sivakoff}, \& {Wong}}]{jennings14}
{Jennings}, Z.~G., {Strader}, J., {Romanowsky}, A.~J., {et~al.} 2014, \aj, 148,
  32

\bibitem[{{Jord{\'a}n} {et~al.}(2004){Jord{\'a}n}, {C{\^o}t{\'e}}, {Ferrarese},
  {Blakeslee}, {Mei}, {Merritt}, {Milosavljevi{\'c}}, {Peng}, {Tonry}, \&
  {West}}]{jordan04}
{Jord{\'a}n}, A., {C{\^o}t{\'e}}, P., {Ferrarese}, L., {et~al.} 2004, \apj,
  613, 279

\bibitem[{{Kashyap} {et~al.}(2010){Kashyap}, {van Dyk}, {Connors}, {Freeman},
  {Siemiginowska}, {Xu}, \& {Zezas}}]{kashyap10}
{Kashyap}, V.~L., {van Dyk}, D.~A., {Connors}, A., {et~al.} 2010, \apj, 719,
  900

\bibitem[{{Kim} {et~al.}(2013){Kim}, {Fabbiano}, {Ivanova}, {Fragos},
  {Jord{\'a}n}, {Sivakoff}, \& {Voss}}]{kim13}
{Kim}, D.-W., {Fabbiano}, G., {Ivanova}, N., {et~al.} 2013, \apj, 764, 98

\bibitem[{{Kim} {et~al.}(2006){Kim}, {Fabbiano}, {Kalogera}, {King},
  {Pellegrini}, {Trinchieri}, {Zepf}, {Zezas}, {Angelini}, {Davies}, \&
  {Gallagher}}]{kim06}
{Kim}, D.-W., {Fabbiano}, G., {Kalogera}, V., {et~al.} 2006, \apj, 652, 1090

\bibitem[{{Kim} {et~al.}(2009){Kim}, {Fabbiano}, {Brassington}, {Fragos},
  {Kalogera}, {Zezas}, {Jord{\'a}n}, {Sivakoff}, {Kundu}, {Zepf}, {Angelini},
  {Davies}, {Gallagher}, {Juett}, {King}, {Pellegrini}, {Sarazin}, \&
  {Trinchieri}}]{kim09}
{Kim}, D.-W., {Fabbiano}, G., {Brassington}, N.~J., {et~al.} 2009, \apj, 703,
  829

\bibitem[{{Liu} {et~al.}(2015){Liu}, {Peng}, {C{\^o}t{\'e}}, {Ferrarese},
  {Jord{\'a}n}, {Mihos}, {Zhang}, {Mu{\~n}oz}, {Puzia}, {Lan{\c c}on}, {Gwyn},
  {Cuillandre}, {Blakeslee}, {Boselli}, {Durrell}, {Duc}, {Guhathakurta},
  {MacArthur}, {Mei}, {S{\'a}nchez-Janssen}, \& {Xu}}]{liu15b}
{Liu}, C., {Peng}, E.~W., {C{\^o}t{\'e}}, P., {et~al.} 2015, \apj, 812, 34

\bibitem[{{Luo} {et~al.}(2013){Luo}, {Fabbiano}, {Strader}, {Kim}, {Brodie},
  {Fragos}, {Gallagher}, {King}, \& {Zezas}}]{luo13}
{Luo}, B., {Fabbiano}, G., {Strader}, J., {et~al.} 2013, \apjs, 204, 14

\bibitem[{{Maccarone} {et~al.}(2005){Maccarone}, {Fender}, \&
  {Tzioumis}}]{maccarone05}
{Maccarone}, T.~J., {Fender}, R.~P., \& {Tzioumis}, A.~K. 2005, \apss, 300, 239

\bibitem[{{Madrid}(2011)}]{madrid11}
{Madrid}, J.~P. 2011, \apjl, 737, L13

\bibitem[{{Madrid} \& {Donzelli}(2013)}]{madrid13}
{Madrid}, J.~P., \& {Donzelli}, C.~J. 2013, \apj, 770, 158

\bibitem[{{Madrid} {et~al.}(2010){Madrid}, {Graham}, {Harris}, {Goudfrooij},
  {Forbes}, {Carter}, {Blakeslee}, {Spitler}, \& {Ferguson}}]{madrid10}
{Madrid}, J.~P., {Graham}, A.~W., {Harris}, W.~E., {et~al.} 2010, \apj, 722,
  1707

\bibitem[{{Maraston} {et~al.}(2004){Maraston}, {Bastian}, {Saglia},
  {Kissler-Patig}, {Schweizer}, \& {Goudfrooij}}]{maraston04}
{Maraston}, C., {Bastian}, N., {Saglia}, R.~P., {et~al.} 2004, \aap, 416, 467

\bibitem[{{Martini} \& {Ho}(2004)}]{martini04}
{Martini}, P., \& {Ho}, L.~C. 2004, \apj, 610, 233

\bibitem[{{Merloni} {et~al.}(2003){Merloni}, {Heinz}, \& {di
  Matteo}}]{merloni03}
{Merloni}, A., {Heinz}, S., \& {di Matteo}, T. 2003, \mnras, 345, 1057

\bibitem[{{Mieske} {et~al.}(2013){Mieske}, {Frank}, {Baumgardt},
  {L{\"u}tzgendorf}, {Neumayer}, \& {Hilker}}]{mieske13}
{Mieske}, S., {Frank}, M.~J., {Baumgardt}, H., {et~al.} 2013, \aap, 558, A14

\bibitem[{{Mieske} {et~al.}(2008{\natexlab{a}}){Mieske}, {Hilker}, {Bomans},
  {Rey}, {Kim}, {Yoon}, \& {Chung}}]{mieske08uv}
{Mieske}, S., {Hilker}, M., {Bomans}, D.~J., {et~al.} 2008{\natexlab{a}}, \aap,
  489, 1023

\bibitem[{{Mieske} {et~al.}(2007{\natexlab{a}}){Mieske}, {Hilker},
  {Jord{\'a}n}, {Infante}, \& {Kissler-Patig}}]{mieske07}
{Mieske}, S., {Hilker}, M., {Jord{\'a}n}, A., {Infante}, L., \&
  {Kissler-Patig}, M. 2007{\natexlab{a}}, \aap, 472, 111

\bibitem[{{Mieske} {et~al.}(2009){Mieske}, {Hilker}, {Misgeld}, {Jord{\'a}n},
  {Infante}, \& {Kissler-Patig}}]{mieske09}
{Mieske}, S., {Hilker}, M., {Misgeld}, I., {et~al.} 2009, \aap, 498, 705

\bibitem[{{Mieske} \& {Kroupa}(2008)}]{mieskekroupa08}
{Mieske}, S., \& {Kroupa}, P. 2008, \apj, 677, 276

\bibitem[{{Mieske} {et~al.}(2007{\natexlab{b}}){Mieske}, {West}, \& {de
  Oliveira}}]{mieske07b}
{Mieske}, S., {West}, M.~J., \& {de Oliveira}, C.~M. 2007{\natexlab{b}}, in
  Groups of Galaxies in the Nearby Universe, ed. I.~{Saviane}, V.~D. {Ivanov},
  \& J.~{Borissova}, 103

\bibitem[{{Mieske} {et~al.}(2008{\natexlab{b}}){Mieske}, {Hilker},
  {Jord{\'a}n}, {Infante}, {Kissler-Patig}, {Rejkuba}, {Richtler},
  {C{\^o}t{\'e}}, {Baumgardt}, {West}, {Ferrarese}, \& {Peng}}]{mieske08}
{Mieske}, S., {Hilker}, M., {Jord{\'a}n}, A., {et~al.} 2008{\natexlab{b}},
  \aap, 487, 921

\bibitem[{{Miller} {et~al.}(2012){Miller}, {Gallo}, {Treu}, \&
  {Woo}}]{miller12}
{Miller}, B., {Gallo}, E., {Treu}, T., \& {Woo}, J.-H. 2012, \apj, 747, 57

\bibitem[{{Miller} {et~al.}(2015){Miller}, {Gallo}, {Greene}, {Kelly}, {Treu},
  {Woo}, \& {Baldassare}}]{miller15}
{Miller}, B.~P., {Gallo}, E., {Greene}, J.~E., {et~al.} 2015, \apj, 799, 98

\bibitem[{{Misgeld} \& {Hilker}(2011)}]{misgeldhilker11}
{Misgeld}, I., \& {Hilker}, M. 2011, \mnras, 414, 3699

\bibitem[{{Misgeld} {et~al.}(2008){Misgeld}, {Mieske}, \& {Hilker}}]{misgeld08}
{Misgeld}, I., {Mieske}, S., \& {Hilker}, M. 2008, \aap, 486, 697

\bibitem[{{Misgeld} {et~al.}(2011){Misgeld}, {Mieske}, {Hilker}, {Richtler},
  {Georgiev}, \& {Schuberth}}]{misgeld11}
{Misgeld}, I., {Mieske}, S., {Hilker}, M., {et~al.} 2011, \aap, 531, A4

\bibitem[{{Nandra} {et~al.}(1997){Nandra}, {George}, {Mushotzky}, {Turner}, \&
  {Yaqoob}}]{nandra97}
{Nandra}, K., {George}, I.~M., {Mushotzky}, R.~F., {Turner}, T.~J., \&
  {Yaqoob}, T. 1997, \apj, 476, 70

\bibitem[{{Norris} \& {Kannappan}(2011)}]{norris11}
{Norris}, M.~A., \& {Kannappan}, S.~J. 2011, \mnras, 414, 739

\bibitem[{{Norris} {et~al.}(2014){Norris}, {Kannappan}, {Forbes}, {Romanowsky},
  {Brodie}, {Faifer}, {Huxor}, \& {Maraston}}]{norris14}
{Norris}, M.~A., {Kannappan}, S.~J., {Forbes}, D.~A., {et~al.} 2014, \mnras,
  443, 1151

\bibitem[{{Paudel} {et~al.}(2010){Paudel}, {Lisker}, \& {Janz}}]{paudel10}
{Paudel}, S., {Lisker}, T., \& {Janz}, J. 2010, \apjl, 724, L64

\bibitem[{{Peng} {et~al.}(2004){Peng}, {Ford}, \& {Freeman}}]{peng04}
{Peng}, E.~W., {Ford}, H.~C., \& {Freeman}, K.~C. 2004, \apjs, 150, 367

\bibitem[{{Penny} {et~al.}(2012){Penny}, {Forbes}, \& {Conselice}}]{penny12}
{Penny}, S.~J., {Forbes}, D.~A., \& {Conselice}, C.~J. 2012, \mnras, 422, 885

\bibitem[{{Penny} {et~al.}(2014){Penny}, {Forbes}, {Strader}, {Usher},
  {Brodie}, \& {Romanowsky}}]{penny14}
{Penny}, S.~J., {Forbes}, D.~A., {Strader}, J., {et~al.} 2014, \mnras, 439,
  3808

\bibitem[{{Peterson}(2001)}]{peterson01}
{Peterson}, B.~M. 2001, in Advanced Lectures on the Starburst-AGN, ed.
  I.~{Aretxaga}, D.~{Kunth}, \& R.~{M{\'u}jica}, 3

\bibitem[{{Phillipps} {et~al.}(2001){Phillipps}, {Drinkwater}, {Gregg}, \&
  {Jones}}]{phillipps01}
{Phillipps}, S., {Drinkwater}, M.~J., {Gregg}, M.~D., \& {Jones}, J.~B. 2001,
  \apj, 560, 201

\bibitem[{{Phillipps} {et~al.}(2013){Phillipps}, {Young}, {Drinkwater},
  {Gregg}, \& {Karick}}]{phillipps13}
{Phillipps}, S., {Young}, A.~J., {Drinkwater}, M.~J., {Gregg}, M.~D., \&
  {Karick}, A. 2013, \mnras, 433, 1444

\bibitem[{{Plotkin} {et~al.}(2012){Plotkin}, {Markoff}, {Kelly}, {K{\"o}rding},
  \& {Anderson}}]{plotkin12}
{Plotkin}, R.~M., {Markoff}, S., {Kelly}, B.~C., {K{\"o}rding}, E., \&
  {Anderson}, S.~F. 2012, \mnras, 419, 267

\bibitem[{{Ptak} {et~al.}(1998){Ptak}, {Yaqoob}, {Mushotzky}, {Serlemitsos}, \&
  {Griffiths}}]{ptak98}
{Ptak}, A., {Yaqoob}, T., {Mushotzky}, R., {Serlemitsos}, P., \& {Griffiths},
  R. 1998, \apjl, 501, L37

\bibitem[{{Reines} {et~al.}(2014){Reines}, {Plotkin}, {Russell}, {Mezcua},
  {Condon}, {Sivakoff}, \& {Johnson}}]{reines14}
{Reines}, A.~E., {Plotkin}, R.~M., {Russell}, T.~D., {et~al.} 2014, \apjl, 787,
  L30

\bibitem[{{Reines} {et~al.}(2011){Reines}, {Sivakoff}, {Johnson}, \&
  {Brogan}}]{reines11}
{Reines}, A.~E., {Sivakoff}, G.~R., {Johnson}, K.~E., \& {Brogan}, C.~L. 2011,
  \nat, 470, 66

\bibitem[{{Rejkuba} {et~al.}(2007){Rejkuba}, {Dubath}, {Minniti}, \&
  {Meylan}}]{rejkuba07}
{Rejkuba}, M., {Dubath}, P., {Minniti}, D., \& {Meylan}, G. 2007, \aap, 469,
  147

\bibitem[{{Remillard} \& {McClintock}(2006)}]{remillard06}
{Remillard}, R.~A., \& {McClintock}, J.~E. 2006, \araa, 44, 49

\bibitem[{{Sandoval} {et~al.}(2015){Sandoval}, {Vo}, {Romanowsky}, {Strader},
  {Choi}, {Jennings}, {Conroy}, {Brodie}, {Foster}, {Villaume}, {Norris},
  {Janz}, \& {Forbes}}]{sandoval15}
{Sandoval}, M.~A., {Vo}, R.~P., {Romanowsky}, A.~J., {et~al.} 2015, \apjl, 808,
  L32

\bibitem[{{Schweizer} \& {Seitzer}(1998)}]{schweizer98}
{Schweizer}, F., \& {Seitzer}, P. 1998, \aj, 116, 2206

\bibitem[{{Seth et al.}(2014)}]{seth14}
{Seth et al.} 2014, \nat, 513, 398

\bibitem[{{Sivakoff} {et~al.}(2007){Sivakoff}, {Jord{\'a}n}, {Sarazin},
  {Blakeslee}, {C{\^o}t{\'e}}, {Ferrarese}, {Juett}, {Mei}, \&
  {Peng}}]{sivakoff07}
{Sivakoff}, G.~R., {Jord{\'a}n}, A., {Sarazin}, C.~L., {et~al.} 2007, \apj,
  660, 1246

\bibitem[{{Spitzer}(1987)}]{spitzer87}
{Spitzer}, L. 1987, {Dynamical evolution of globular clusters}

\bibitem[{{Stark} {et~al.}(1992){Stark}, {Gammie}, {Wilson}, {Bally}, {Linke},
  {Heiles}, \& {Hurwitz}}]{stark92}
{Stark}, A.~A., {Gammie}, C.~F., {Wilson}, R.~W., {et~al.} 1992, \apjs, 79, 77

\bibitem[{{Strader} {et~al.}(2012{\natexlab{a}}){Strader}, {Chomiuk},
  {Maccarone}, {Miller-Jones}, \& {Seth}}]{strader12b}
{Strader}, J., {Chomiuk}, L., {Maccarone}, T.~J., {Miller-Jones}, J.~C.~A., \&
  {Seth}, A.~C. 2012{\natexlab{a}}, \nat, 490, 71

\bibitem[{{Strader} {et~al.}(2012{\natexlab{b}}){Strader}, {Chomiuk},
  {Maccarone}, {Miller-Jones}, {Seth}, {Heinke}, \& {Sivakoff}}]{strader12}
{Strader}, J., {Chomiuk}, L., {Maccarone}, T.~J., {et~al.} 2012{\natexlab{b}},
  \apjl, 750, L27

\bibitem[{{Strader} {et~al.}(2013){Strader}, {Seth}, {Forbes}, {Fabbiano},
  {Romanowsky}, {Brodie}, {Conroy}, {Caldwell}, {Pota}, {Usher}, \&
  {Arnold}}]{strader13}
{Strader}, J., {Seth}, A.~C., {Forbes}, D.~A., {et~al.} 2013, \apjl, 775, L6

\bibitem[{{Thornton} {et~al.}(2009){Thornton}, {Barth}, {Ho}, \&
  {Greene}}]{thornton09}
{Thornton}, C.~E., {Barth}, A.~J., {Ho}, L.~C., \& {Greene}, J.~E. 2009, \apj,
  705, 1196

\bibitem[{{Vaughan} {et~al.}(2003){Vaughan}, {Edelson}, {Warwick}, \&
  {Uttley}}]{vaughan03}
{Vaughan}, S., {Edelson}, R., {Warwick}, R.~S., \& {Uttley}, P. 2003, \mnras,
  345, 1271

\bibitem[{{Wehner} \& {Harris}(2007)}]{wehner07}
{Wehner}, E.~M.~H., \& {Harris}, W.~E. 2007, \apjl, 668, L35

\bibitem[{{Young} {et~al.}(2012){Young}, {Brandt}, {Xue}, {Paolillo},
  {Alexander}, {Bauer}, {Lehmer}, {Luo}, {Shemmer}, {Schneider}, \&
  {Vignali}}]{young12}
{Young}, M., {Brandt}, W.~N., {Xue}, Y.~Q., {et~al.} 2012, \apj, 748, 124

\bibitem[{{Zhang} {et~al.}(2015){Zhang}, {Peng}, {C{\^o}t{\'e}}, {Liu},
  {Ferrarese}, {Cuillandre}, {Caldwell}, {Gwyn}, {Jord{\'a}n}, {Lan{\c c}on},
  \& {Li}}]{zhang15}
{Zhang}, H.-X., {Peng}, E.~W., {C{\^o}t{\'e}}, P., {et~al.} 2015, \apj, 802, 30

\end{thebibliography}

\appendix
\section{Chandra Observations Used}\label{sec:obs}
The following numbers refer to the Observation Identifiers (``ObsIDs") of unique \textit{Chandra} datasets used in this work.
\begin{enumerate}
\item \textit{Fornax cluster:} 14527, 239, 2942, 319, 320, 4168, 4169, 4170, 4171, 4172, 4173, 4174, 4176, 4177, 4742, 624, 9526, 9530, 9798, 9799
\item \textit{Coma cluster:} 555, 556, 1086, 1112, 1113, 1114, 9714, 10672, 13993, 13994, 13995, 13996, 14406, 14410, 14411, 14415
\item \textit{Perseus cluster:} 502, 503, 1513, 3209, 4289, 4946, 4947, 4948, 4949, 4950, 4951, 4952, 4953, 6139, 6145, 6146, 11713, 11714, 11715, 11716, 12025, 12033, 12036, 12037
\item \textit{Virgo cluster:} 321,   322,   352,   517,   539,   785,  2016,  2068,  2072, 2707,  3717,  5826,  5827,  5828,  6186,  7210,  7211,  7212, 7864,  8046,  8047,  8050,  8057,  8058,  8063,  8064,  8071, 8074,  8078,  8079,  8090,  8094,  8095,  8098,  8103,  8107, 8109,  8110,  8127,  8131,  8182,  8507,  8581, 10241, 11274, 11783, 12888, 12889, 12975, 12976, 12978, 13985, 14328, 14358, 14359, 15178, 15179, 15180, 15757, 15758, 15759, 15760, 16033, 16260, 16261, 16585, 16586, 16587, 16590, 16591, 16592, 16593
\item \textit{Hydra I cluster:} 2220
\item \textit{Centaurus cluster:} 504, 505, 4190, 4191, 4954, 4955, 5310, 8179, 16223, 16224, 16225, 16534, 16607, 16608, 16609, 16610
\item \textit{NGC 1132 fossil group:} 3576, 801
\item \textit{NGC 1023 ``poor" group:} 7131
\item \textit{Hickson compact groups 22 and 90:} 8172, 905
\item \textit{NGC 3115 galaxy:} 2040, 11268, 12095, 13817, 13819, 13820, 13821, 13822, 14383, 14384, 14419
\item \textit{NGC 7252 galaxy:} 2980
\item \textit{M104 (Sombrero) galaxy:} 1586, 9532, 9533
\item \textit{Centaurus A (NGC 5128) galaxy:} 10722, 10723, 10724, 10725, 10726, 11846, 11847, 12155, 12156, 13303, 13304, 15294, 15295, 16276, 2978, 316, 3965, 7797, 7798, 7799, 7800, 8489, 8490, 962
\end{enumerate}

\section{Borderline Detections}\label{sec:borderline}
The following five UCDs present 0.5-7.0 keV (full band) net counts which are consistent with zero (i.e., the error bars are larger than the central value; see \autoref{tab:xraydet}): F-18, gregg25, gregg26, gregg45, and M85-HCC1. Since \texttt{wavdetect} did find these sources to be significant (taking into account the spatial distribution of source and background region counts), we further scrutinized these UCDs' source and background regions in the individual contributing observations. Some of the UCDs required a redefinition of the background region due to contamination from a nearby bright source.

For F-18, gregg25, gregg45, and M85-HCC1, the total number of counts in the source region is greater than the number of local background counts by a factor of $\sim2$. The net counts are also significantly greater than a fluctuation expected from the local background with Poisson probability $1\%$ (between 1.5 to 2.75 times greater). Therefore, we consider these four UCDs to be bona fide X-ray-detected UCDs (although we note that they are all significantly off-axis). 

For gregg26, the total counts and local background counts are 31 and 25 counts respectively, and so the margin of significance is very low. The net counts are also significantly less than a fluctuation expected from the local background with Poisson probability $1\%$: 5.98 compared to 12 counts. Visual inspection of its postage stamp in \autoref{fig:stamps} does suggest an over-density of counts above that seen in the local background. Since \texttt{wavdetect} found a significant detection for gregg26 taking into account the spatial distribution of the counts in the source and background apertures, we will continue to consider gregg26 an X-ray-detected UCD in what follows. Although including gregg26 does not affect the main results of our study, we do caution that its identification as an X-ray-emitting UCD is not completely secure. 

\section{Comments on Perseus and NGC 3115}\label{sec:perseus3115}
In both Perseus and NGC 3115, we have extraordinarily high exposure times, up to $\sim1.4$ Msec and $\sim1.1$ Msec respectively, but no unambiguous detections (among the 84 Perseus UCDs and 31 NGC 3115 UCDs). There are $\sim3$ localized over-densities found by \texttt{wavdetect} in Perseus, each of which is within $\sim1.5$ arcsec of a unique UCD. However, none of those appear to be genuine point sources and seem instead to arise from a knot of soft-X-ray-emitting hot gas. To further scrutinize those candidates, we created images in an even harder energy band that is presumably further devoid of emission from the thermally-emitting hot gas: 4-7 keV. With most of the thermal emission from the gas gone in these ``harder-band" images and the lack of X-ray point sources near UCDs, we confirmed that there is no evidence in favor of X-ray-emitting UCDs in Perseus, including the extremely massive Perseus-UCD13. 

\citet{jennings14} found that UCD20 in NGC 3115 is an X-ray emitter. We do find an X-ray point source near UCD20, but it is offset from the optical position of UCD20 by $\sim3.76$ arcsec. Given that the 90\% uncertainty on the absolute astrometry of the Chandra dataset is $\sim0.6$ arcsec, that offset would correspond to a $\gtrsim6\sigma$ positional error, which means that a match is unlikely. We compared the positions of the point sources in our merged image to known objects listed in the Chandra Source Catalog \citep{evans10} and found excellent agreement. Therefore, we do not consider UCD20 to be an X-ray emitter. 

\section{Optical Properties and X-ray Upper Limits for Non-X-ray-detected UCDs}\label{sec:nondetprop}
Here, we describe our derivation of upper limits for non-X-ray-detected UCDs. Just as for the X-ray-detected UCDs, we first extracted and combined the individual contributing observations' source and background region spectra for each non-X-ray-detected UCD using \texttt{srcflux} and \texttt{combine\_spectra}. This extraction+combination step will also create the Auxiliary Response File (ARF) which gives the product of effective area (cm$^2$) and quantum efficiency (counts/photon) as a function of energy for a detector, and the ARF-weighted Redistribution Matrix File (RMF) which gives the response and finite resolution elements of the ACIS detectors used in the individual observations. These empirical exposure-weighted ARFs and RMFs yield more careful estimates of the count rate compared to the default, perfectly on-axis ACIS ARF and RMF assumed by tools such as WebPIMMS.

We next compute the scaled background counts, $C_B$, by multiplying the total counts observed in the background region by the ratio of source-to-background region areas, which is a standard practice to account for the fact that the background region is typically much larger than the source region. Then, we construct the inverse of the cumulative distribution function of the Poisson distribution characterized by mean $C_B$, and compute from that the total number of counts corresponding to the $99\%$ tail of the function. We used the Python module \texttt{scipy.stats.poisson.ppf} for this task. The upper limit on net counts, $\mathcal{U}$, is thus defined as the total number of counts expected from such a fluctuation minus $C_B$. We consider the amplitude of this fluctuation above the local background, with Poisson probability $1\%$, to be the maximum possible number of net counts contributed by a source while still making that source indistinguishable from the local background (i.e., just barely non-detected at our adopted detection threshold).

As a sanity check, we verified that the net counts actually measured in a source region were lower than the $\mathcal{U}$ net counts calculated above assuming a Poisson fluctuation in the local background. This criterion was satisfied in nearly all cases. In the remaining few cases, we visually inspected the source and background regions to make sure that a detection was not hidden in the noise (which is what would be suggested if the actual net counts $>\mathcal{U}$). These cases typically occur in very high exposure time observations where there can be significant inhomogeneities in the local background (e.g., due to a knot of thermally-emitting hot gas). In some cases, we also had to redefine the background and/or source regions to avoid contamination or background over-subtraction due to a nearby point source. The crucial point is that we carefully checked all upper limits to ensure that there are no ``missed" detections.

To go from units of net counts to flux, we first divide the $\mathcal{U}$ net counts by the summed exposure time, giving a net count rate. Then, we divide that net count rate by the product of the effective area and quantum efficiency which results in a photon flux.\footnote{The value of the ARF at a given energy yields the product of the effective area and quantum efficiency. In low-count cases, it is prudent to use the ARF value at the ARF-weighted mean energy.} Finally, we multiply that photon flux by the ARF-weighted mean energy per photon giving the conventional energy flux in units of [erg s$^{-1}$ cm$^{-2}$]. Using the luminosity-distance formula with the assumed distance to the UCD's host system, we arrive at a luminosity in units of [erg s$^{-1}$]. 

\begin{sidewaystable}
\caption{Optical and X-ray Properties of Non-X-ray-detected UCDs\label{tab:nondetprop}}
\tiny 
\resizebox{\linewidth}{!}{
\begin{tabular}{cccccccccccccccccccc}
\tableline
UCD & System & RA & Dec & $D$ & $M_V$ & log(M/M$_{\odot}$) & r$_{\rm hl}$ & $v_{\rm rad}$ & $\sigma$ & M$_{dyn}$/L$_V$ & EXP & $\theta$ & R & C$_{S+B}$ & C$_{B}$ & $\mathcal{U}$ & F$_F$ & L$_F$ & Ref \\
(1) & (2) & (3) & (4) & (5) & (6) & (7) & (8) & (9) & (10) & (11) & (12) & (13) & (14) & (15) & (16) & (17) & (18) & (19) & (20) \\ \tableline
f3\_1 & NGC 1023 & 38.41833 & 37.80639 & 12.25 & $-$ & $-$ & $-$ & $-$ & $-$ & $-$ & $-$ & $-$ & $-$ & $-$ & $-$ & $-$ & $-$ & $-$ & Mieske+07b \\
f3\_71 & NGC 1023 & 38.51042 & 38.60917 & 12.25 & $-$ & $-$ & $-$ & $-$ & $-$ & $-$ & $-$ & $-$ & $-$ & $-$ & $-$ & $-$ & $-$ & $-$ & Mieske+07b \\
f2\_43 & NGC 1023 & 38.55042 & 38.95472 & 12.25 & $-$ & $-$ & $-$ & $-$ & $-$ & $-$ & $-$ & $-$ & $-$ & $-$ & $-$ & $-$ & $-$ & $-$ & Mieske+07b \\
f3\_52 & NGC 1023 & 38.69333 & 38.19083 & 12.25 & $-$ & $-$ & $-$ & $-$ & $-$ & $-$ & $-$ & $-$ & $-$ & $-$ & $-$ & $-$ & $-$ & $-$ & Mieske+07b \\
f37 & NGC 1023 & 38.74333 & 37.88222 & 12.25 & $-$ & $-$ & $-$ & $-$ & $-$ & $-$ & $-$ & $-$ & $-$ & $-$ & $-$ & $-$ & $-$ & $-$ & Mieske+07b \\
f25 & NGC 1023 & 38.75833 & 38.78278 & 12.25 & $-$ & $-$ & $-$ & $-$ & $-$ & $-$ & $-$ & $-$ & $-$ & $-$ & $-$ & $-$ & $-$ & $-$ & Mieske+07b \\
f3\_77 & NGC 1023 & 38.94375 & 38.51389 & 12.25 & $-$ & $-$ & $-$ & $-$ & $-$ & $-$ & $-$ & $-$ & $-$ & $-$ & $-$ & $-$ & $-$ & $-$ & Mieske+07b \\
f2\_45 & NGC 1023 & 38.94917 & 39.05611 & 12.25 & $-$ & $-$ & $-$ & $-$ & $-$ & $-$ & 4 & 9.4 & 20.4 & 0.0 & 1.3 & 3.7 & $<-$13.7 & <38.6 & Mieske+07b \\
f2\_51 & NGC 1023 & 39.02833 & 39.30056 & 12.25 & $-$ & $-$ & $-$ & $-$ & $-$ & $-$ & $-$ & $-$ & $-$ & $-$ & $-$ & $-$ & $-$ & $-$ & Mieske+07b \\
f2\_56 & NGC 1023 & 39.26333 & 39.43778 & 12.25 & $-$ & $-$ & $-$ & $-$ & $-$ & $-$ & $-$ & $-$ & $-$ & $-$ & $-$ & $-$ & $-$ & $-$ & Mieske+07b \\
f2\_53 & NGC 1023 & 39.39417 & 39.20555 & 12.25 & $-$ & $-$ & $-$ & $-$ & $-$ & $-$ & $-$ & $-$ & $-$ & $-$ & $-$ & $-$ & $-$ & $-$ & Mieske+07b \\
f1\_20 & NGC 1023 & 39.89792 & 39.43583 & 12.25 & $-$ & $-$ & $-$ & $-$ & $-$ & $-$ & $-$ & $-$ & $-$ & $-$ & $-$ & $-$ & $-$ & $-$ & Mieske+07b \\
f46 & NGC 1023 & 40.07375 & 37.88083 & 12.25 & $-$ & $-$ & $-$ & $-$ & $-$ & $-$ & $-$ & $-$ & $-$ & $-$ & $-$ & $-$ & $-$ & $-$ & Mieske+07b \\
f49 & NGC 1023 & 40.40750 & 37.79861 & 12.25 & $-$ & $-$ & $-$ & $-$ & $-$ & $-$ & $-$ & $-$ & $-$ & $-$ & $-$ & $-$ & $-$ & $-$ & Mieske+07b \\
f4\_23 & NGC 1023 & 40.68250 & 38.09750 & 12.25 & $-$ & $-$ & $-$ & $-$ & $-$ & $-$ & $-$ & $-$ & $-$ & $-$ & $-$ & $-$ & $-$ & $-$ & Mieske+07b \\
UCD1 & NGC 1132 & 43.21250 & $-$1.27194 & 99.5 & $-$ & $-$ & 77.1 & 7158.0 & $-$ & $-$ & 53 & 0.9 & 1.7 & 4.0 & 3.6 & 5.4 & $<-$14.7 & <39.4 & Madrid+13 \\
UCD6 & NGC 1132 & 43.21250 & $-$1.26833 & 99.5 & $-$ & $-$ & 9.9 & 7147.0 & $-$ & $-$ & 53 & 1.0 & 1.8 & 1.0 & 2.4 & 4.6 & $<-$14.7 & <39.4 & Madrid+13 \\
UCD4 & NGC 1132 & 43.21667 & $-$1.26194 & 99.5 & $-$ & $-$ & 16.7 & 6627.0 & $-$ & $-$ & 53 & 1.2 & 1.8 & 0.0 & 1.5 & 3.5 & $<-$14.8 & <39.2 & Madrid+13 \\
UCD2 & NGC 1132 & 43.22500 & $-$1.29417 & 99.5 & $-$ & $-$ & 13.3 & 6834.0 & $-$ & $-$ & 53 & 0.8 & 1.7 & 0.0 & 0.9 & 3.1 & $<-$14.9 & <39.2 & Madrid+13 \\
UCD5 & NGC 1132 & 43.22917 & $-$1.28861 & 99.5 & $-$ & $-$ & 13.0 & 6949.0 & $-$ & $-$ & 53 & 0.6 & 1.7 & 2.0 & 1.2 & 2.8 & $<-$14.9 & <39.1 & Madrid+13 \\
UCD3 & NGC 1132 & 43.24583 & $-$1.24833 & 99.5 & $-$ & $-$ & 19.8 & 7082.0 & $-$ & $-$ & 53 & 2.4 & 2.7 & 1.0 & 1.0 & 3.0 & $<-$14.9 & <39.1 & Madrid+13 \\
HCG22\_UCD011 & HCG 22+90 & 45.87537 & $-$15.63583 & 33.1 & $-$ & $-$ & $-$ & 2766.0 & $-$ & $-$ & 31 & 1.9 & 2.2 & 2.0 & 0.1 & 0.9 & $<-$15.2 & <37.9 & DaRocha+11 \\
HCG22\_UCD010 & HCG 22+90 & 45.88608 & $-$15.64139 & 33.1 & $-$ & $-$ & $-$ & 2755.0 & $-$ & $-$ & 31 & 1.3 & 1.8 & 0.0 & 0.3 & 1.7 & $<-$14.9 & <38.2 & DaRocha+11 \\
HCG22\_UCD007 & HCG 22+90 & 45.89267 & $-$15.61756 & 33.1 & $-$ & $-$ & $-$ & 2548.0 & $-$ & $-$ & 31 & 1.7 & 2.1 & 1.0 & 0.1 & 0.9 & $<-$15.2 & <37.9 & DaRocha+11 \\
HCG22\_UCD001 & HCG 22+90 & 45.89358 & $-$15.61272 & 33.1 & $-$ & $-$ & $-$ & 2834.0 & $-$ & $-$ & 31 & 2.0 & 2.3 & 1.0 & 0.1 & 0.9 & $<-$15.2 & <37.9 & DaRocha+11 \\
HCG22\_UCD008 & HCG 22+90 & 45.89600 & $-$15.62525 & 33.1 & $-$ & $-$ & $-$ & 2944.0 & $-$ & $-$ & 31 & 1.3 & 1.8 & 0.0 & 0.3 & 1.7 & $<-$14.9 & <38.2 & DaRocha+11 \\
HCG22\_UCD002 & HCG 22+90 & 45.90100 & $-$15.62022 & 33.1 & $-$ & $-$ & $-$ & 2552.0 & $-$ & $-$ & 31 & 1.4 & 1.9 & 0.0 & 0.6 & 2.4 & $<-$14.8 & <38.3 & DaRocha+11 \\
HCG22\_UCD012 & HCG 22+90 & 45.90133 & $-$15.62106 & 33.1 & $-$ & $-$ & $-$ & 2549.0 & $-$ & $-$ & 31 & 1.3 & 1.9 & 1.0 & 0.3 & 1.7 & $<-$14.9 & <38.2 & DaRocha+11 \\
HCG22\_UCD016 & HCG 22+90 & 45.90863 & $-$15.62183 & 33.1 & $-$ & $-$ & $-$ & 2487.0 & $-$ & $-$ & 31 & 1.2 & 1.9 & 0.0 & 0.5 & 2.5 & $<-$14.8 & <38.4 & DaRocha+11 \\
HCG22\_UCD009 & HCG 22+90 & 45.91479 & $-$15.62717 & 33.1 & $-$ & $-$ & $-$ & 2211.0 & $-$ & $-$ & 31 & 1.0 & 1.8 & 1.0 & 0.2 & 1.8 & $<-$14.9 & <38.2 & DaRocha+11 \\
HCG22\_UCD015 & HCG 22+90 & 45.91742 & $-$15.61692 & 33.1 & $-$ & $-$ & $-$ & 2789.0 & $-$ & $-$ & 31 & 1.6 & 2.1 & 0.0 & 0.3 & 1.7 & $<-$14.9 & <38.2 & DaRocha+11 \\
HCG22\_UCD014 & HCG 22+90 & 45.91896 & $-$15.60519 & 33.1 & $-$ & $-$ & $-$ & 2905.0 & $-$ & $-$ & 31 & 2.3 & 2.7 & 0.0 & 0.5 & 2.5 & $<-$14.8 & <38.3 & DaRocha+11 \\
HCG22\_UCD006 & HCG 22+90 & 45.91975 & $-$15.62375 & 33.1 & $-$ & $-$ & $-$ & 2654.0 & $-$ & $-$ & 31 & 1.3 & 1.9 & 0.0 & 0.1 & 0.9 & $<-$15.2 & <37.9 & DaRocha+11 \\
HCG22\_UCD005 & HCG 22+90 & 45.91996 & $-$15.61586 & 33.1 & $-$ & $-$ & $-$ & 2665.0 & $-$ & $-$ & 31 & 1.7 & 2.1 & 1.0 & 0.3 & 1.7 & $<-$14.9 & <38.2 & DaRocha+11 \\
HCG22\_UCD003 & HCG 22+90 & 45.93308 & $-$15.62028 & 33.1 & $-$ & $-$ & $-$ & 2652.0 & $-$ & $-$ & 31 & 2.0 & 2.3 & 1.0 & 0.1 & 0.9 & $<-$15.2 & <37.9 & DaRocha+11 \\
HCG22\_UCD004 & HCG 22+90 & 45.93592 & $-$15.62872 & 33.1 & $-$ & $-$ & $-$ & 2668.0 & $-$ & $-$ & 31 & 1.8 & 2.3 & 0.0 & 0.3 & 1.7 & $<-$14.9 & <38.2 & DaRocha+11 \\
HCG22\_UCD013 & HCG 22+90 & 45.93721 & $-$15.61519 & 33.1 & $-$ & $-$ & $-$ & 2646.0 & $-$ & $-$ & 31 & 2.3 & 2.7 & 1.0 & 0.3 & 1.7 & $<-$14.9 & <38.2 & DaRocha+11 \\
UCD60 & Perseus & 49.86792 & 41.57500 & 71.1 & $-$ & $-$ & 49.0 & 3735.0 & $-$ & $-$ & 1420 & 4.5 & 6.9 & 4215.0 & 4289.2 & 152.8 & $<-$14.6 & <39.2 & Penny+12 \\
UCD65 & Perseus & 49.88879 & 41.57619 & 71.1 & $-$ & $-$ & 13.6 & 7681.0 & $-$ & $-$ & 1368 & 4.4 & 6.7 & 4952.0 & 5118.8 & 167.2 & $<-$14.5 & <39.3 & Penny+12 \\
UCD1 & Perseus & 49.90683 & 41.49550 & 71.1 & $-$ & $-$ & 11.7 & 5717.0 & $-$ & $-$ & 1494 & 2.3 & 3.6 & 14554.0 & 14664.8 & 282.2 & $<-$14.4 & <39.4 & Penny+12 \\
UCD2 & Perseus & 49.91579 & 41.50478 & 71.1 & $-$ & $-$ & 26.9 & 5687.0 & $-$ & $-$ & 1494 & 2.7 & 3.7 & 21241.0 & 20989.0 & 338.0 & $<-$14.3 & <39.5 & Penny+12 \\
UCD7 & Perseus & 49.92096 & 41.53642 & 71.1 & $-$ & $-$ & 12.5 & 5272.0 & $-$ & $-$ & 1494 & 2.4 & 4.3 & 13676.0 & 13683.0 & 273.0 & $<-$14.4 & <39.4 & Penny+12 \\
UCD9 & Perseus & 49.92333 & 41.53553 & 71.1 & $-$ & $-$ & 14.4 & 9955.0 & $-$ & $-$ & 1494 & 2.4 & 4.3 & 14695.0 & 14592.9 & 282.1 & $<-$14.4 & <39.4 & Penny+12 \\
UCD13 & Perseus & 49.93804 & 41.53500 & 71.1 & $-$12.81 & 8.36172783602 & 85.0 & 5292.0 & 38.0 & 19.85 & 1457 & 2.1 & 4.2 & 15919.0 & 16516.0 & 300.0 & $<-$14.3 & <39.5 & Penny+12 \\
UCD17 & Perseus & 49.95233 & 41.52161 & 71.1 & $-$ & $-$ & 18.4 & 5286.0 & $-$ & $-$ & 1457 & 2.5 & 4.3 & 28921.0 & 28362.5 & 392.5 & $<-$14.2 & <39.6 & Penny+12 \\
UCD19 & Perseus & 49.95817 & 41.52267 & 71.1 & $-$ & $-$ & 13.1 & 5680.0 & $-$ & $-$ & 1457 & 2.6 & 4.4 & 26781.0 & 27052.7 & 383.3 & $<-$14.2 & <39.6 & Penny+12 \\
UCD21 & Perseus & 49.95854 & 41.53581 & 71.1 & $-$ & $-$ & 31.2 & 5668.0 & $-$ & $-$ & 1402 & 2.8 & 4.8 & 20547.0 & 20440.0 & 333.0 & $<-$14.3 & <39.5 & Penny+12 \\
UCD23 & Perseus & 49.95896 & 41.48731 & 71.1 & $-$ & $-$ & 11.5 & 5693.0 & $-$ & $-$ & 1297 & 3.5 & 5.0 & 6668.0 & 6445.0 & 187.0 & $<-$14.5 & <39.3 & Penny+12 \\
UCD25 & Perseus & 49.96429 & 41.50481 & 71.1 & $-$ & $-$ & 11.5 & 4750.0 & $-$ & $-$ & 1297 & 3.1 & 4.4 & 17854.0 & 17893.0 & 312.0 & $<-$14.3 & <39.5 & Penny+12 \\
UCD29 & Perseus & 49.96996 & 41.50444 & 71.1 & $-$ & $-$ & 22.4 & 5250.0 & $-$ & $-$ & 1212 & 3.1 & 4.3 & 9949.0 & 10014.8 & 233.2 & $<-$14.3 & <39.4 & Penny+12 \\
UCD40 & Perseus & 49.99475 & 41.49864 & 71.1 & $-$ & $-$ & 13.9 & 5632.0 & $-$ & $-$ & 1139 & 3.9 & 5.6 & 4446.0 & 4580.3 & 157.7 & $<-$14.5 & <39.3 & Penny+12 \\
\end{tabular}}
\textbf{Notes:} Upper limits were derived only for spectroscopically-confirmed UCDs, with the exception of NGC 3115 (which has $\sim1.1$ Msec of imaging) and NGC 1023 (which has no spectroscopically-confirmed UCDs). The UCDs are sorted by right ascension and declination so that all UCDs from the same host system are found in the same part of this table. Col (1): UCD name. Col (2): UCD host system. Col (3): UCD right ascension [deg]. Col (4): UCD declination [deg]. Col (5): Host system distance [Mpc]. Col (6): Absolute magnitude in $V$-band. Col (7): $\log_{10}$ of (dynamical mass)/$M_{\odot}$. Col (8): Half-light radius [pc]. Col (9): Radial velocity [km s$^{-1}$]. Col (10): Global stellar velocity dispersion [km s$^{-1}$]. Col (11): Dynamical mass-to-light ratio in the $V$-band. Col (12): Total \textit{Chandra} exposure time [ksec]. Col (13): Exposure time-weighted average off-axis angle of all observations used to derive upper limit [arcmin]. Col (14): Exposure time-weighted average radius of the circular apertures used to enclose 90\% of the PSF in the individual observations [arcsec]. Col (15): Total counts in the source region [counts]. Col (16): Scaled background counts in the background region [counts]. Col (17): Number of counts expected from a fluctuation in the local background with Poisson probability $1\%$ [counts]. Col (18): $\log_{10}$ of 0.5-7.0 keV flux upper limit based on $\mathcal{U}$ counts [erg s$^{-1}$ cm$^{-2}$]. Col (19): $\log_{10}$ of 0.5-7.0 keV luminosity upper limit based on $mathcal{U}$ counts [erg s$^{-1}$]. Col (20): Primary reference paper for UCD's optical properties. \textit{The full version of this table is available online.}
\end{sidewaystable}

\end{document}